\documentclass[fleqn,usenatbib,useAMS]{mnras}

\usepackage{graphicx}	
\usepackage{amsmath}	
\usepackage{multicol}        
\usepackage{bm}		
\usepackage{pdflscape}	
\usepackage{latexsym}
\usepackage[authoryear]{natbib}
\usepackage{bm,nicefrac}
\usepackage{subfig}

\usepackage[T1]{fontenc}
\usepackage{ae,aecompl}

\usepackage{newtxtext,newtxmath}

\def\deg{\ifmmode^\circ\else$^\circ$\fi}

\def\Q{\ifmmode\mathcal{Q}\else$\mathcal{Q}$\fi}
\def\Mach{\ifmmode\mathcal{M}\else$\mathcal{M}$\fi}

\pdfoutput=1

\title[New insights in the Galactic `Snake' nebula]
{Galactic `Snake' IRDC G11.11$-$0.12: a site of multiple hub-filament systems and colliding filamentary clouds}
\author[L.~K. Dewangan et al.]{L.~K. Dewangan$^{1}$\thanks{E-mail: lokeshd@prl.res.in}, 
N.~K. Bhadari$^{1}$, 
A.~K. Maity$^{1,2}$,
C. Eswaraiah$^{3}$,
Saurabh Sharma$^{4}$, 
\newauthor 
and O.~R. Jadhav$^{1,2}$
 \\ 
$^{1}$Physical Research Laboratory, Navrangpura, Ahmedabad - 380 009, India.\\
$^{2}$Indian Institute of Technology Gandhinagar Palaj, Gandhinagar 382355, India.\\ 
$^{3}$Indian Institute of Science Education and Research (IISER) Tirupati, Rami Reddy Nagar, Karakambadi Road,
Mangalam (P.O.), Tirupati 517 507, India.\\ 
$^{4}$Aryabhatta Research Institute of Observational Sciences, Manora Peak, Nainital 263002, India.
} 
\begin{document}

\date{ }

\pagerange{\pageref{firstpage}--\pageref{lastpage}} \pubyear{2020}

\maketitle

\label{firstpage}

\begin{abstract}
To probe star formation processes, we present a multi-scale and multi-wavelength investigation of the `Snake' nebula/infrared dark cloud G11.11$-$0.12 (hereafter, G11; length $\sim$27 pc). 
{\it Spitzer} images hint at the presence of sub-filaments (in absorption), and reveal four infrared-dark hub-filament 
system (HFS) candidates (extent $<$ 6 pc) toward G11, where massive clumps ($>$ 500 {\it M$_{\odot}$}) and protostars are identified. 
The $^{13}$CO(2--1), C$^{18}$O(2--1), and NH$_{3}$(1,1) line data reveal a noticeable velocity oscillation toward G11, 
as well as its left part (or part-A) around V$_{lsr}$ of 31.5 km s$^{-1}$, and its right part (or part-B) around V$_{lsr}$ of 29.5 km s$^{-1}$. 
The common zone of these cloud components is investigated toward the center's G11 housing one HFS. 
Each cloud component hosts two sub-filaments. 
In comparison to part-A, more ATLASGAL clumps are observed toward part-B. The \emph{JWST} near-infrared images discover one 
infrared-dark HFS candidate (extent $\sim$0.55 pc) around the massive protostar G11P1 (i.e., G11P1-HFS). 
Hence, the infrared observations reveal multiple infrared-dark HFS candidates at multi-scale in G11.
The ALMA 1.16 mm continuum map shows multiple finger-like features (extent $\sim$3500--10000 AU) surrounding a dusty envelope-like feature (extent $\sim$18000 AU) toward 
the central hub of G11P1-HFS. Signatures of forming massive stars are found toward the center of the envelope-like feature. 
The ALMA H$^{13}$CO$^{+}$ line data show two cloud components with a velocity separation of $\sim$2 km s$^{-1}$ toward G11P1.
Overall, the collision process, the ``fray and fragment'' mechanism, and the ``global non-isotropic collapse'' scenario seem to be operational in G11. 
\end{abstract}
%
\section{Introduction}
\label{sec:intro}
The past decade has witnessed a significant improvement in understanding the formation processes of massive OB-stars 
({\it M} $\gtrsim$ 8 {\it M$_{\odot}$}). However, the underlying physical mechanisms for the mass transfer (or mass accumulation) from parsec-scale clumps to cores in massive star formation (MSF) research are still unknown \citep{Motte+2018,rosen20}. Such processes can be studied in infrared dark clouds (IRDCs) \citep[or elongated dust filaments as absorption feature at infrared (IR) wavelengths; e.g.,][]{rathborne06}, and  dust/molecular filaments in emission (at sub-millimeter (sub-mm) wavelengths) hosting hub-filament systems \citep[HFSs;][]{myers09}, massive young stellar objects (MYSOs), and Class II methanol masers at 6.7 GHz \citep[e.g.,][]{walsh98,minier01}. However, the complex mechanisms involved in MSF and the origin and internal structures of filaments are not fully understood \citep[e.g.,][]{tan14,hacar22}. 
In this context, the popular scenarios are the cloud-cloud collision \citep[CCC;][]{habe92,fukui21} or converging flows \citep{ballesteros99,vazqez07,heitsch08,beuther20}, the ``fray and fragment'' \citep{tafalla15}, and the clump-fed \citep{bonnell_2001,bonnell_2004,vazquez_2009,padoan20}, which have received considerable attention recently. 
The former two scenarios are connected with the collision, while the latter one is related to the inflow material from very large-scales. 
The clump-fed scenario predicts that low-mass fragments grow into massive stars within HFSs/clouds \citep[e.g.,][]{Motte+2018}. 
To observationally assess the highlighted scenarios, one needs to carefully study the initial conditions that lead to the birth of young massive stars and
observed structures (i.e., filaments and HFSs). It demands to observationally infer the velocity field and the spatial morphology toward star-forming structures at various physical scales. 

The present work focuses on the `Snake' nebula or G11.11$-$0.12 (hereafter, G11), which is one of the well explored filamentary IRDCs using multi-wavelength data sets \citep[including molecular line data, continuum maps, and polarization observations; see][for more details]{chen22}. The IRDC G11 is prominently evident in the {\it Spitzer} 8.0 $\mu$m image as presented in Figure~\ref{fig1}a. 
Previously, using a distance of $\sim$3.6 kpc to G11 \citep{carey98,carey00}, the length and the mass of G11 were reported to be $\sim$30 pc and $\sim$10$^{5}$ {\it M$_{\odot}$} \citep{pillai06,henning10,kainulainen13}, respectively. In Figure~\ref{fig1}a, the {\it Spitzer} 8.0 $\mu$m image is also overlaid with the APEX Telescope Large Area Survey of the Galaxy \citep[ATLASGAL;][]{schuller09} clumps at 870 $\mu$m. The reliable distance (i.e., $\sim$2.92 kpc) of these clumps (see Table~\ref{tab2xx} for physical parameters) has been determined by \citet{urquhart18}, who utilized the H\,{\sc i} analysis, maser parallax, and spectroscopic measurements to resolve the distance ambiguity for clumps (see their paper for more details). Note that the previously reported distance differs from the distance of $\sim$2.92 kpc, which is adopted in this present work.  The presence of dense gas associated with clumps/cores, young stellar objects (YSOs), water maser, and 
sign-posts of MSF (i.e., 6.7 GHz Class II methanol maser emission) has been investigated toward G11, suggesting the ongoing early phases of star formation activity \citep[e.g.,][]{carey98,carey00,johnstone03,pillai06,pillai15,pillai19,chen10,henning10,gomez11,bhavya13,kainulainen13,rosero14,shipman14,wang14,ragan15,schneider15,
lin17,tafoya21,chen22,ngoc23}. 

Several previous works also focused on a compact dust continuum source associated with the point-like mid-IR (MIR) emission, the water maser, and the 6.7 GHz methanol maser, favoring the presence of a massive proto-stellar candidate G11.11$-$0.12P1 \citep[or G11P1; see][for more details]{rosero14}. 
Using the {\it Spitzer} 8.0 $\mu$m image, \citet{pillai15} suggested the presence of a HFS toward the central part of the IRDC G11. 
Based on the outcomes derived from a variety of recent observations, this promising IRDC can allow us to investigate ongoing 
physical processes and the structure of a young, massive, and star-forming cloud, both of which have not been thoroughly studied. 

In this relation, we have carefully examined several published and unpublished multi-wavelength and multi-scale data sets. 
To explore the internal sub-structures of the filamentary cloud and/or the proposed HFS toward the IRDC G11, 
we carefully examined the {\it Spitzer} 8.0 $\mu$m image. 
The present study also examines the velocity structure or spatial-kinematic structure of the molecular gas in G11 using the $^{13}$CO, C$^{18}$O, and NH$_{3}$(1, 1) line data, allowing us to assess the presence and absence of different velocity components. In the direction of the massive proto-stellar candidate G11P1, we employed the 
Atacama Large Millimeter/submillimeter Array (ALMA) 1.16 mm continuum map and the ALMA H$^{13}$CO$^{+}$(3--2) line data to examine its inner environment ($<$ 20000~AU). 
High-resolution and high-sensitivity near-IR (NIR) observations from the James 
Webb Space Telescope (\emph{JWST}) are also employed toward the clump c7 (or G11P1) and the clump c2 \citep[or G11P6;][and see Table~\ref{tab2xx} in this paper]{wang14}, and are used to examine the dust and gaseous structures below 20000~AU scale \citep[e.g.,][]{reiter22,dewangan2023new}. 

This paper is organized as follows. 
Section~\ref{sec:obser} provides the details of adopted published and unpublished observational data 
sets in this paper. 
Observational outcomes derived using multi-wavelength and multi-scale data sets are presented in Section~\ref{sec:data}. 
It includes the investigation of sub-filaments, multiple HFSs, embedded protostars, dust clumps/cores, 
and velocity structures at multi-scale. 
The implications of our observed results are discussed in Section~\ref{sec:disc}.  
Finally, Section~\ref{sec:conc} presents the main findings.
\section{Data sets and analysis}
\label{sec:obser}
\begin{figure*}
\includegraphics[width=\textwidth]{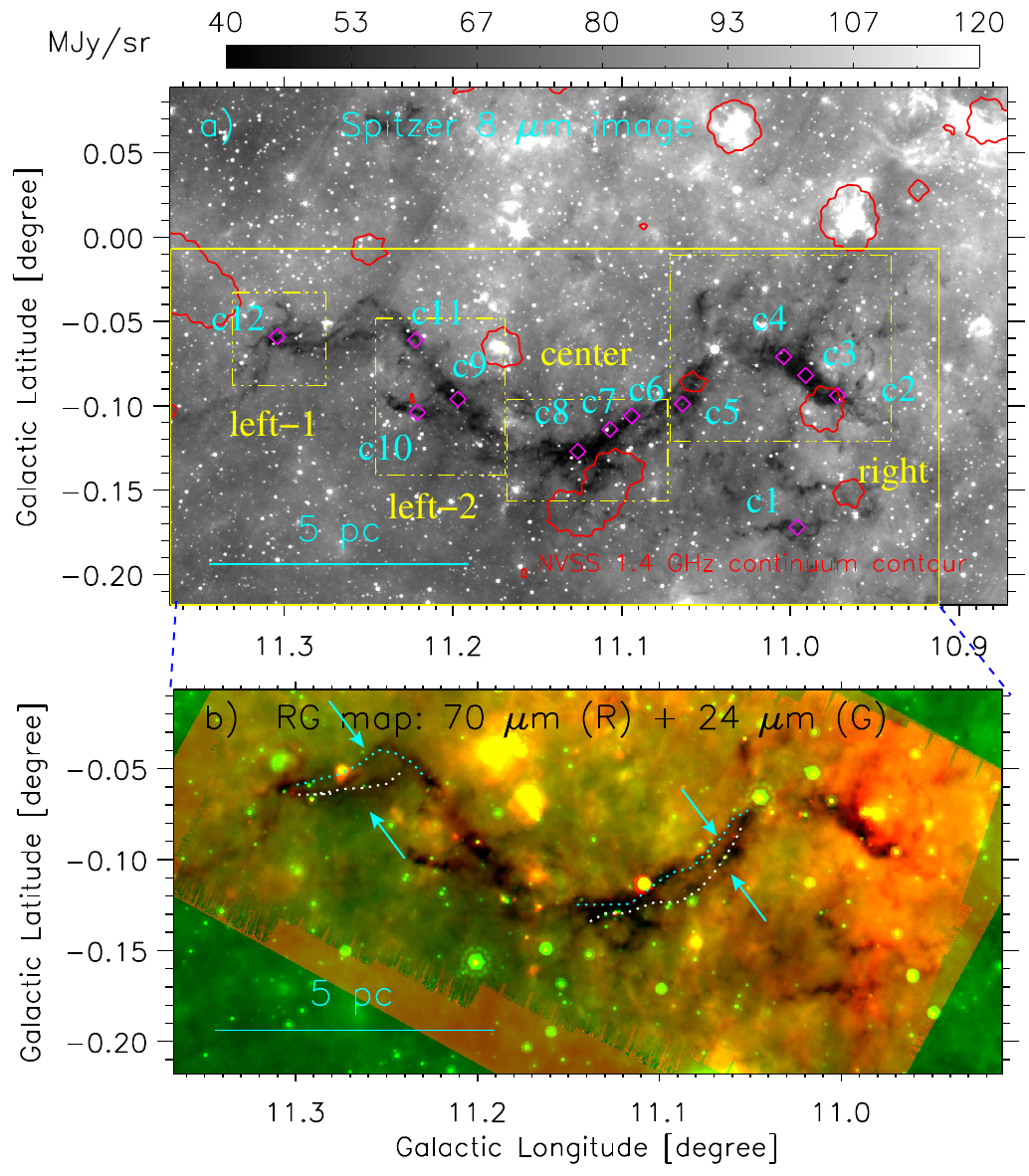}
\caption{a) The panel shows the selected target 
area (i.e., $\sim$29$'$.76 $\times$ 18$'$.36; central coordinates: {\it l} = 11$\degr$.119; {\it b} = $-$0$\degr$.0647) hosting the IRDC G11 
using the {\it Spitzer} 8.0 $\mu$m image. The NVSS 1.4 GHz continuum emission contour (in red) at 10$\sigma$ (where 1$\sigma$ $\sim$0.45 mJy beam$^{-1}$) 
and the positions of the ATLASGAL dust continuum clumps at 870 $\mu$m \citep{urquhart18} are also superposed on 
the image (see magenta diamonds). The solid box encompasses the area presented in Figure~\ref{fig1}b. 
b) Two-color composite map made using the {\it Herschel} 70 $\mu$m image (in red) and the {\it Spitzer} 24 $\mu$m image (in green). 
Arrows highlight the presence of sub-filaments toward the filamentary IRDC G11 (see also two dotted curves). 
Each panel has a scale bar that corresponds to 5 pc (at a distance of 2.92 kpc).} 
\label{fig1}
\end{figure*}

In this work, various observational data sets were obtained for an area of $\sim$29$'$.76 $\times$ 18$'$.36 (central coordinates: {\it l} = 11$\degr$.119; {\it b} = $-$0$\degr$.0647; see Figure~\ref{fig1}a) hosting the IRDC G11. 
Different surveys have been employed to obtain multi-wavelength data sets, 
and these surveys are listed in Table~\ref{tab1}. 

The existing high-resolution ALMA continuum and line data of G11P1 were collected in this work. 
We examined the primary-beam corrected ALMA continuum map at 1.16 mm (1$\sigma$ $\sim$185 $\mu$Jy beam$^{-1}$) and H$^{13}$CO$^{+}$(3--2) line data (velocity resolution $\sim$0.56 km s$^{-1}$) toward G11P1, which were downloaded from the ALMA science archive. The quasar J1924-2914 was utilized for the flux calibration and the bandpass calibrator, while the quasar J1832-2039 was utilized for phase calibration \citep[see also][for details]{lopez21,sanhueza21}. 

We utilized the level-3 science ready \emph{JWST} Near-Infrared Camera \citep[NIRCam;][]{2005SPIE.5904....1R,2012SPIE.8442E..2NB} images of SNAKE-FIELD-1/clump~c7/G11P1 and SNAKE-FIELD-2/clump~c7/G11P6 (Proposal~ID: 1182; Proposal~PI: Young, Erick~T) from the Mikulski Archive for Space Telescopes (MAST) archive. 
One can find more details of \emph{JWST} performance in \citet{rigby2023}. 

The $^{13}$CO(2--1) and C$^{18}$O(2--1) line data cubes were obtained from the SEDIGISM survey, which was carried out using the Atacama Pathfinder EXperiment (APEX) 12m telescope. 
To improve sensitivities, we smoothed the SEDIGISM $^{13}$CO(2--1) and C$^{18}$O(2--1) line data cubes \citep[pixel-scale $\sim$9\rlap.{$''$}5; rms $\sim$0.8--1.0~K;][]{schuller17,schuller21} using a symmetrical Gaussian with a full-width at half-maximum (FWHM) of 3 pixels, 
resulting a angular resolution of $\sim$41\rlap.{$''$}4. The reliable distances and velocities of the ATLASGAL 870 $\mu$m dust continuum clumps distributed toward our selected target area were collected from the published work of \citet{urquhart18} (see Table~\ref{tab2xx}). 

We employed the GLIMPSE-I Spring '07 highly reliable catalog to obtain photometric magnitudes of point-like sources 
at {\it Spitzer} 3.6--8.0 $\mu$m bands. 
Furthermore, we examined photometric magnitudes of sources at {\it Spitzer} 24 $\mu$m obtained from the publicly available MIPSGAL catalog \citep[e.g.,][]{gutermuth15}.  
\begin{table*}
\setlength{\tabcolsep}{0.16in}
\centering
\caption{Table lists the ID, galactic coordinates, 870 $\mu$m peak flux density ($P_{870}$), 
870 $\mu$m integrated flux density ($S_{870}$), radial velocity ($V_\mathrm{lsr}$), distance, clump effective radius ($R_\mathrm{c}$), 
dust temperature ($T_\mathrm{d}$), and clump mass ($M_\mathrm{clump}$) of 12 ATLASGAL 870 $\mu$m dust continuum clumps 
from \citet{urquhart18} in our selected target area (see diamonds in Figure~\ref{fig1}a). Last column contains names of sub-region where the dust clumps are spatially associated (see Figure~\ref{fig1}a). Clumps associated with outflow lobes \citep{yang22} are highlighted by superscript ``$\dagger$''.}
\label{tab2xx}
\begin{tabular}{lcccccccccccccc} 
\hline 										    	        			      
  ID         &   Longitude 	   &      Latitude    &  $P_{870}$  & $S_{870}$  & $V_\mathrm{lsr}$     & distance& $R_\mathrm{c}$&$T_{d}$& $M_\mathrm{clump}$&sub-region\\ 
             &   [degree]	   &    [degree]      &   (Jy beam$^{-1}$) & (Jy)       & (km s$^{-1}$) &  (kpc)  & (pc)          & (K)   & ($M_\odot$)& \\ 
\hline 
c1   &	10.996  &  $-$0.172 & 0.49 &   1.99 & 28.3 &   2.92 &	 0.14 &   13.1  &    185  & --  \\
c2   &	10.972  &  $-$0.094 & 1.35 &   9.11 & 30.0 &   2.92 &	 0.42 &   11.6  &   1052  & right  \\
c3$^{\dagger}$   &	10.991  &  $-$0.082 & 1.22 &  15.06 & 29.7 &   2.92 &	 0.86 &   11.9  &   1660  & right  \\
c4$^{\dagger}$   &	11.004  &  $-$0.071 & 0.84 &   7.24 & 26.7 &   2.92 &	 0.35 &   10.5  &   1014  & right  \\
c5   &	11.064  &  $-$0.099 & 0.85 &  10.11 & 29.3 &   2.92 &	 0.49 &   11.1  &   1270  & right  \\
c6   &	11.094  &  $-$0.106 & 0.64 &  14.92 & 29.4 &   2.92 &	 0.40 &   15.6  &   1030  & center  \\
c7$^{\dagger}$   &	11.107  &  $-$0.114 & 1.76 &  15.00 & 29.8 &   2.92 &	 0.62 &   15.8  &   1014  & center  \\
c8$^{\dagger}$   &	11.126  &  $-$0.127 & 1.11 &  14.76 & 30.0 &   2.92 &	 0.71 &   11.3  &   1795  & center  \\
c9   &	11.197  &  $-$0.096 & 0.57 &   5.59 & 31.7 &   2.92 &	 0.25 &   12.9  &    532  & left-2   \\
c10  &	11.221  &  $-$0.104 & 0.48 &   2.07 & 29.6 &   2.92 &	 0.14 &   10.9  &    270  &  left-2  \\
c11  &	11.222  &  $-$0.061 & 0.47 &   2.17 & 31.8 &   2.92 &	 0.14 &   12.3  &    225  &  left-2  \\
c12  &	11.304  &  $-$0.059 & 0.63 &   5.74 & 31.6 &   2.92 &	 0.23 &   12.9  &    547  & left-1  \\
\hline          		
\end{tabular}			
\end{table*}

%
\begin{table*}
\scriptsize
\setlength{\tabcolsep}{0.18in}
\centering
\caption{Different surveys utilized in this paper.}
\label{tab1}
\begin{tabular}{lcccr}
\hline 
  Survey/facility  &  Wavelength/      &  Resolution        &  Reference \\   
    &  Frequency/line(s)       &   ($\arcsec$)        &   \\   
\hline
\hline 
NRAO VLA Sky Survey (NVSS)       & 1.4 GHz  & $\sim$45 & \citet{NVSS}\\
Radio Ammonia Mid-Plane Survey (RAMPS) & NH$_{3}$(1, 1)  & $\sim$34.7 & \citet{hogge18}\\
ALMA data    & 1.16 mm, H$^{13}$CO$^{+}$(3--2)  & $\sim$0.34 $\times$ 0.27 & project \#2017.1.00101.S; PI: Sanhueza, Patricio\\
Structure, Excitation and Dynamics of the Inner Galactic Interstellar Medium (SEDIGISM) &  $^{13}$CO, C$^{18}$O(2--1) & $\sim$30        &\citet{schuller17}\\
ATLASGAL                 &870 $\mu$m                     & $\sim$19.2        &\citet{schuller09}\\
{\it Herschel} Infrared Galactic Plane Survey (Hi-GAL)                              &70--500 $\mu$m                     & $\sim$6--37         &\citet{molinari10}\\
{\it Spitzer} MIPS Inner Galactic Plane Survey (MIPSGAL)                                         &24 $\mu$m                     & $\sim$6         &\citet{carey05}\\ 
{\it Spitzer} Galactic Legacy Infrared Mid-Plane Survey Extraordinaire (GLIMPSE)       & 8.0  $\mu$m                   & $\sim$2           &\citet{benjamin03}\\
\emph{JWST} ERO NIRCam Long Wavelength (LW) F356W, F444W imaging facility & 3.563, 4.421 $\mu$m                   
& $\sim$0.17          &\citet{2005SPIE.5904....1R,2012SPIE.8442E..2NB}\\ 
\emph{JWST} ERO NIRCam Short Wavelength (SW) F200W imaging facility  & 1.99 $\mu$m                   
& $\sim$0.07          &\citet{2005SPIE.5904....1R,2012SPIE.8442E..2NB}\\ 
\hline          
\end{tabular}
\end{table*}
%
\section{Results}
\label{sec:data}
\subsection{Distribution of ATLASGAL dust clumps toward G11}
\label{sec0}
Figure~\ref{fig1}a presents an elongated filamentary structure in absorption, where 12 ATLASGAL dust 
clumps at 870 $\mu$m (c1--c12) are distributed. To infer the distribution of the ionized emission toward G11, 
the NVSS 1.4 GHz continuum emission contour is also overlaid on the {\it Spitzer} 8.0 $\mu$m image (see Figure~\ref{fig1}a). 

The ATLASGAL dust clumps show ranges of the dust temperature, mass, and radial velocity to be [10.5, 15.8]~K, 
[185, 1795]~{\it M$_{\odot}$}, and [26.7, 31.8]~km s$^{-1}$, respectively (see Table~\ref{tab2xx}). 
These mass values are determined at a distance of 2.92 kpc \citep[see][for more details]{urquhart18}. 
Using the SEDIGISM $^{13}$CO(2--1) and C$^{18}$O(2--1) line data, \citet{yang22} studied the $^{13}$CO 
outflows toward the ATLASGAL clumps, and tabulated the red and blue wing velocity components of the $^{13}$CO(2--1) emission. 
They reported outflow lobes toward 4 out of 12 ATLASGAL clumps (i.e., c3, c4, c7, c8; see Table~\ref{tab2xx}). 
In the direction of the clump c4, \citet{yang22} detected the blue wing (at V$_{lsr}$ $\sim$[26.2, 26.5] km s$^{-1}$) 
and the red wing (at V$_{lsr}$ $\sim$[30.8, 32.2] km s$^{-1}$) of the outflow. 
However, in the case of other three clumps (i.e., c3, c7, and c8), we find only the detection of the red wing velocity component. 
According to \citet{yang22}, the red wing velocity component toward the clumps c3, c7, and c8 is [31.5, 33.0], [30.8, 32.8], 
and [32.2, 33.5] km s$^{-1}$, respectively. In the direction of the clumps c3 and c7 (or G11P1), molecular outflows were also reported in the literature \citep[e.g.,][]{wang14,li20,tafoya21}.

Based on a visual inspection, we have marked four sub-regions (i.e., left-1, left-2, center, 
and right; see dot-dashed boxes in Figure~\ref{fig1}a) toward the IRDC G11. 
The ATLASGAL clumps distributed toward the edges and the central parts of the IRDC G11 are the massive ones ($>$ 500 {\it M$_{\odot}$}; see clumps c2, c8, and c12 in Figure~\ref{fig1}a). 
The most massive clump (i.e., c8) is located at the ``center'' sub-region. 
Interestingly, we find that the massive clump c7, also located in the ``center'' sub-region, hosts the massive proto-stellar candidate G11P1, 
which is associated with the water maser, the 6.7 GHz methanol maser, and a string of four unresolved 
VLA radio continuum sources \citep[e.g.,][]{rosero14}. 
The NVSS radio continuum emission is traced toward two sub-regions, ``center'' and ``right'' (see a red contour in Figure~\ref{fig1}a). 
Considering the radial velocities toward the ATLASGAL clumps \citep[from][]{urquhart18}, the clumps c9, c11, and c12 are traced with a velocity range of [31, 32]~km s$^{-1}$, 
while seven clumps c2--c8 are depicted with a velocity range of [29, 30]~km s$^{-1}$ (see Table~\ref{tab2xx}). 
This particular result shows the presence of the variation in the radial velocity (or velocity difference $\sim$2 km s$^{-1}$) toward the cloud G11. 
Considering this outcome, a detailed analysis of the molecular gas is presented in Section~\ref{sec2}.
\begin{figure*}
\includegraphics[width=\textwidth]{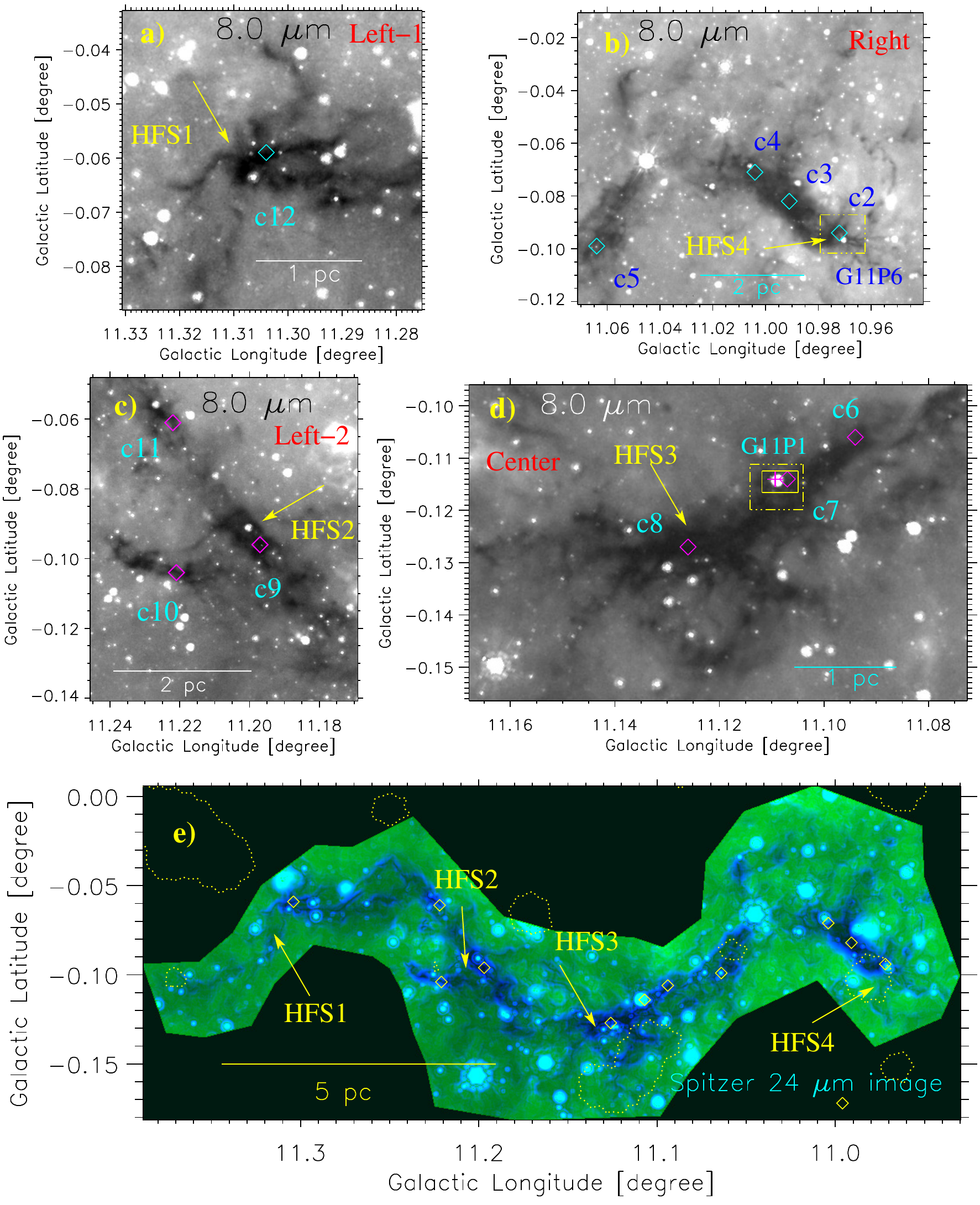}
\caption{a--d) Zoomed-in views of four small parts of the IRDC G11 (i.e., Left-1, Right, Left-2, and Center, respectively) using the {\it Spitzer} 8.0 $\mu$m image (see dot-dashed boxes in Figure~\ref{fig1}a). 
e) Overlay of the NVSS 1.4 GHz continuum emission contour (in yellow) and the positions of the ATLASGAL clumps (see yellow diamonds) 
on a two-color composite map made using the {\it Spitzer} 24 $\mu$m image (in green) and the {\it ``sobel''} processed {\it Spitzer} 24 $\mu$m image (in blue). 
The dot-dashed box in panel ``b'' encompasses the area presented in Figure~\ref{fig3lc}c, 
while the area shown in Figure~\ref{fig3lc}a is indicated by the dot-dashed box in panel ``d''. 
In the panel ``d'', the solid box (in yellow) is shown as a zoomed-in view in Figure~\ref{fig10}, and  a cross indicates the position of the 6.7 GHz methanol maser.
In each panel, diamonds show the positions of the ATLASGAL clumps and a scale bar represents a spatial scale at a distance of 2.92 kpc.}
\label{fig3}
\end{figure*}
\subsection{Embedded internal structures in G11}
\label{sec1}
\subsubsection{Sub-filaments and multiple HFS candidates}
\label{subsec:sstemp}
We have performed a careful visual inspection of the {\it Spitzer} 8.0 $\mu$m image, which displays 
the presence of sub-filaments in absorption toward the filamentary IRDC G11 (see Figure~\ref{fig1}a). 
In order to further explore the sub-filaments, we present a two-color composite map 
(70 $\mu$m (red) and 24 $\mu$m (green) images) of the IRDC G11 in Figure~\ref{fig1}b. 
These IR images also reveal two sub-filaments (in absorption) toward some parts of the IRDC, which are highlighted by arrows and dotted curves in Figure~\ref{fig1}b. 
However, these sub-filaments are not prominently seen toward the entire IRDC G11. 

Additionally, the clumps c2, c8, c9, and c12 seem 
to be surrounded by several parsec-scale filaments in absorption, suggesting the existence of multiple IR-dark HFS candidates. 
To examine zoomed-in views of areas around these four clumps, 
Figures~\ref{fig3}a,~\ref{fig3}b,~\ref{fig3}c, and~\ref{fig3}d present the {\it Spitzer} 8.0 $\mu$m images toward 
the sub-regions left-1, right, left-2, and center, respectively (see dot-dashed boxes in Figure~\ref{fig1}a). 
The positions of the ATLASGAL clumps are also marked in each panel of Figure~\ref{fig3}. 
At least one massive clump ($>$ 500 {\it M$_{\odot}$}; see Table~\ref{tab2xx}) appears 
to be seen toward the central hub of each proposed IR-dark HFS candidate (extent $<$ 6 pc).
 These configurations are treated as small-scale HFS candidates in this work.
Figure~\ref{fig3}e displays a two-color composite map made using the {\it Spitzer} 24 $\mu$m image (in green) 
and the IDL-based routine {\it ``sobel''} \citep{sobel19683x3} processed {\it Spitzer} 24 $\mu$m image (blue), 
which is also overlaid with the ATLASGAL clumps and the NVSS radio continuum contour.  
The {\it ``sobel''} operator is used for edge detection, and identifies the boundaries or edges by computing 
the gradient of image intensity at each pixel within the target image. 
Hence, the composite map seems to reveal four IR-dark HFS candidates (i.e., HFS1--4) and sub-filaments, which are new outcomes in the IRDC G11.

Overall, to validate the existence of sub-filaments and IR-dark HFS candidates, we require new continuum 
and line observations (resolution $<$ 2$''$) of the entire IRDC at sub-mm and millimeter wavelengths. 
\subsubsection{{\it Herschel} maps and embedded protostars}
\label{subsec:temp}
\begin{figure*}
\includegraphics[width=\textwidth]{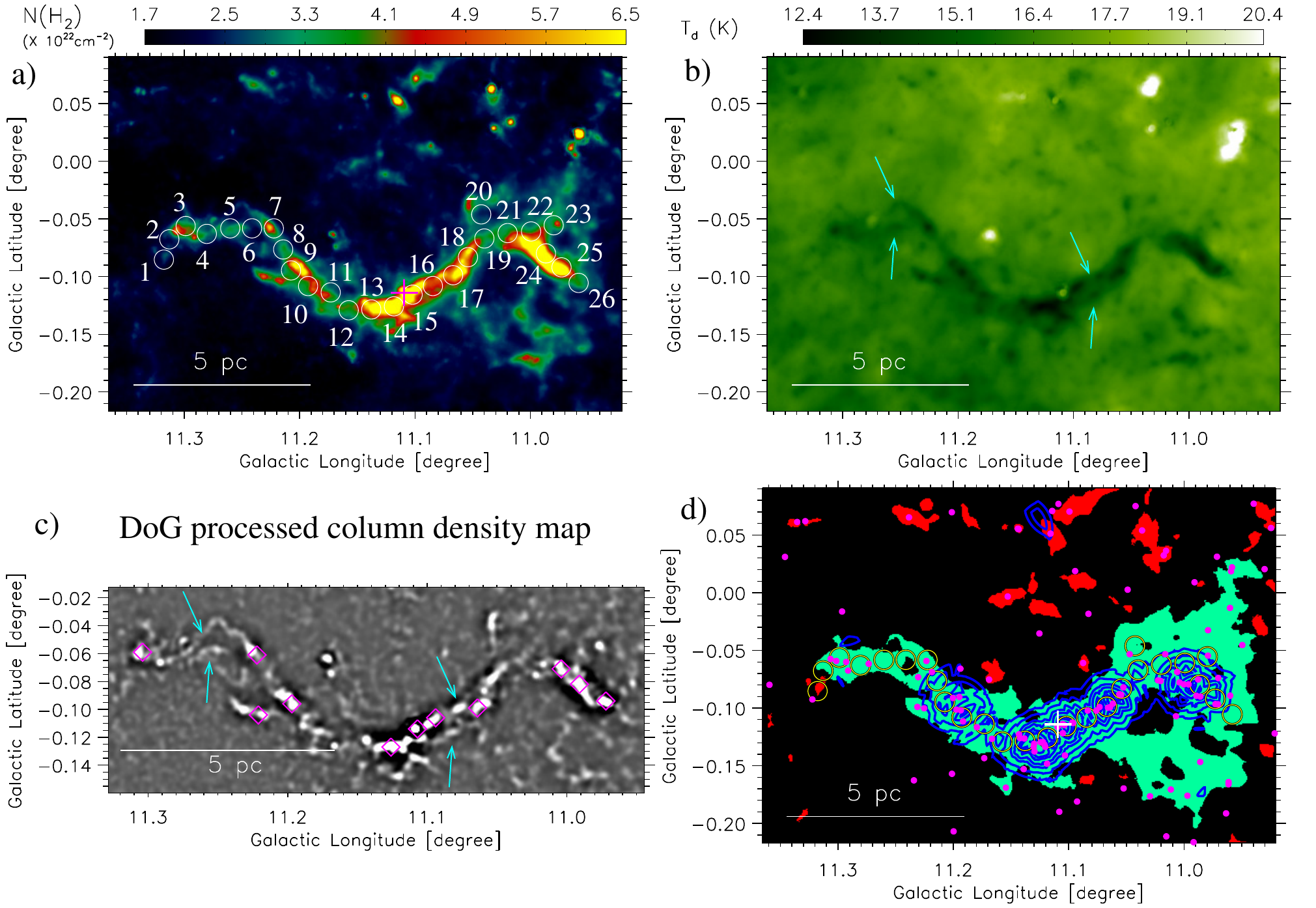}
\caption{a) {\it Herschel} column density (N(H$_{2}$)) map (resolution $\sim$13\rlap.{$''$}5) overlaid with 26 small circular (radius $\sim$30$''$) regions, where some physical parameters are extracted (see Figure~\ref{fig6}). 
b) {\it Herschel} temperature map (resolution $\sim$13\rlap.{$''$}5). 
c) The panel displays the {\it Herschel} column density map processed with the {\it Edge-DoG} algorithm in the direction of the `Snake' nebula, 
which is overlaid with the positions of the ATLASGAL clumps (see diamonds).  
d) The panel presents the filled column density contour (at 2.45 $\times$ 10$^{22}$ cm$^{-2}$) map, which is overlaid with the surface density 
contours (in blue) of YSOs (at (0.7, 1, 1.5, 2, 3, 4, 7) YSOs pc$^{-1}$; where 1$\sigma$ $\sim$0.57 YSOs pc$^{-1}$), the positions of 
the selected protostars (see small filled circles), and the selected circular regions (see open circles and also panel ``a''). 
In panels ``b'' and ``c'', arrows highlight the presence of sub-filaments toward the filamentary IRDC G11.
In panels ``a'' and ``d'', a cross indicates the position of the 6.7 GHz methanol maser. 
A scale bar in each panel is the same as in Figure~\ref{fig1}a.} 
\label{fig4}
\end{figure*}

In Figures~\ref{fig4}a and~\ref{fig4}b, we display the {\it Herschel} column density ($N(\mathrm H_2)$) and 
temperature ($T_\mathrm{d}$) maps (resolution $\sim$13\rlap.{$''$}5) of our selected target area containing the IRDC G11, respectively. 
The steps needed to obtain these maps using the {\it hires} are described in \citet{getsf_2022} \citep[see also Section~2 in][for more details]{dewangan23}. 

The column density map shows the distribution of $H_2$ column density toward the IRDC G11, where 26 circular regions (radius $\sim$30$''$) 
are selected arbitrarily along its spine to further study the variations of $N(\mathrm H_2)$ and $T_\mathrm{d}$. 
We have computed the averaged values of $N(\mathrm H_2)$ and $T_\mathrm{d}$ towards each 
circular region (see Figures~\ref{fig6}a and~\ref{fig6}b). The center and right sub-regions are seen with high 
column densities ($>$ 7 $\times$ 10$^{22}$ cm$^{-2}$). 
The dust temperature distribution ($T_\mathrm{d}$ range $\sim$[12.5, 18]~K) along the filamentary structure can be examined in Figure~\ref{fig4}b.  
Relatively higher dust temperatures are evident toward the massive proto-stellar candidate G11P1. 
Figure~\ref{fig4}c shows the {\it Herschel} column density map processed through the {\it Edge-DoG} algorithm, which utilizes the method of Difference of Gaussians filters \citep[e.g.,][]{Assirati2014}. Arrows highlight elongated sub-filaments, which are seen in emission (see also arrows in Figure~\ref{fig4}b). 

In Figure~\ref{fig4}d, we present the filled column density contour (at 2.45 $\times$ 10$^{22}$ cm$^{-2}$) map, showing the elongated structure.
We have computed the mass of this strucutre to be $\sim$5.5 $\times$ 10$^{4}$ {\it M$_{\odot}$} using the equation, $M_{clump} = \mu_{H_2} m_H A_{pix} \Sigma N(H_2)$, where $\mu_{H_2}$ is the mean molecular weight per hydrogen molecule (i.e., 2.8), $A_{pix}$ is the area subtended by one pixel (i.e., 3$''$.2/pixel), and $\Sigma N(\mathrm H_2)$ is the total column density \citep[see also][]{dewangan17a}. 

In order to identify embedded and young protostars (i.e., Class~I~protostars and flat-spectrum~soruces) toward our target area containing G11, we employed 
the [3.6] $-$ [24]/[3.6] color-magnitude diagram \citep[e.g.,][]{guieu10,rebull11,dewangan15} 
and the [4.5]$-$[5.8] vs [3.6]$-$[4.5] color-color diagram \citep[e.g.,][]{hartmann05,getman07,dewangan15}. 
These plots are not shown in this paper, and one can find more details about these schemes in \citet{dewangan15}. 
In this paper, we have selected a total of 123 protostars \citep[mean age $<$ 1 Myr;][]{evans09}. 
In Figure~\ref{fig4}d, the positions of these selected young protostars are shown by filled circles (in magenta). 
To examine the groups of these selected protostars, the surface density map of 123 protostars is generated in a similar 
way as carried out by \citet{dewangan15} \citep[see also][]{dewangan22l}. 
In this procedure, we used a 5$''$ grid and 6 nearest-neighbor (NN) at a distance of 2.92 kpc.
Figure~\ref{fig4}d also displays the surface density contours (in blue) of the protostars with the levels of 0.7, 1, 1.5, 2, 3, 4, and 7 YSOs pc$^{-2}$ (where 1$\sigma$ = 0.57 YSOs pc$^{-2}$). 
This analysis confirms early stages of ongoing star formation activities in the entire IRDC.
 
\begin{figure*}
\includegraphics[width=15.3 cm]{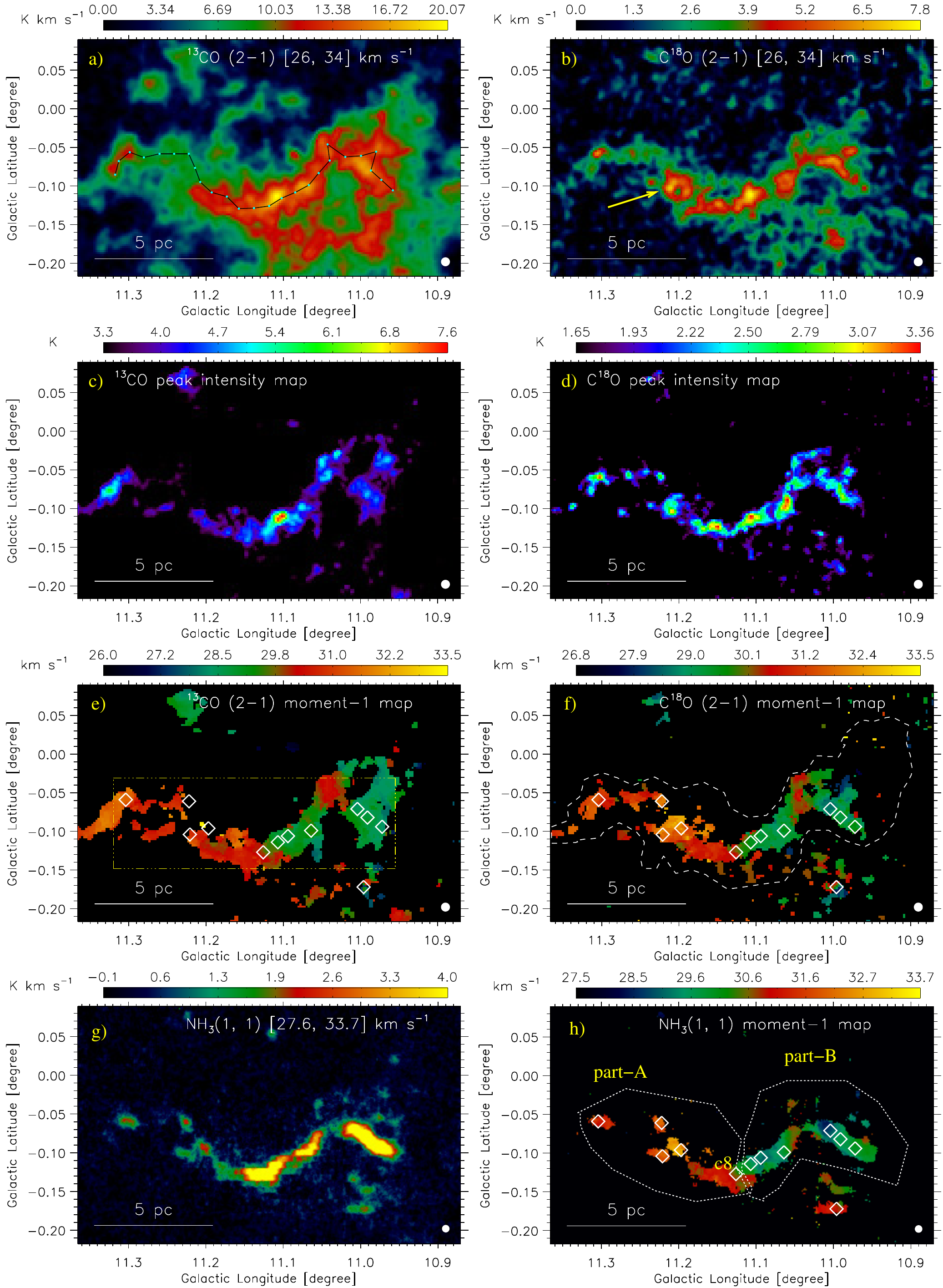}
\caption{a) SEDIGISM $^{13}$CO($J$ = 2--1) map of integrated intensity (moment-0) toward the cloud G11 (see Figure~\ref{fig1}a). 
The molecular emission is integrated over a velocity range from 26 to 34 km s$^{-1}$. 
A curve is marked in the molecular map, where small filled circles show the central locations of 26 small circular regions marked in Figure~\ref{fig4}a.
b) SEDIGISM C$^{18}$O($J$ = 2--1) moment-0 map at [26, 34] km s$^{-1}$. An arrow highlights the circular pattern detected in the molecular map. 
c) SEDIGISM $^{13}$CO($J$ = 2--1) peak intensity map. 
d) SEDIGISM C$^{18}$O($J$ = 2--1) peak intensity map. 
e) SEDIGISM $^{13}$CO($J$ = 2--1) moment-1 map. 
f) SEDIGISM C$^{18}$O($J$ = 2--1) moment-1 map. 
The molecular cloud mass estimation is performed for the region indicated by the dashed curve. 
g) RAMPS NH$_{3}$(1, 1) moment-0 map at [27.6, 33.7] km s$^{-1}$
h) RAMPS NH$_{3}$(1, 1) moment-1 map. Two groups (i.e., ``part-A'' and ``part-B'') are indicated by dotted curves in the panel. 
In panels ``e'', ``f'', and ``h'', diamonds represent the positions of the ATLASGAL clumps. 
In each panel, the filled circle represents the beam size. 
A scale bar in each panel is the same as in Figure~\ref{fig1}a.}
\label{fig5}
\end{figure*}
\subsection{SEDIGISM and RAMPS molecular line data}
\label{sec2}
Our selected target IRDC is well covered in the SEDIGISM and RAMPS surveys. Hence, in order to study the spatial distribution of molecular gas and velocity structure, 
we explored the $^{13}$CO ($J$ = 2--1), C$^{18}$O ($J$ = 2--1), and NH$_{3}$(1, 1) line data toward the IRDC G11. 
\subsubsection{Spatial and velocity structure of molecular gas in G11}
\label{subsec2}
Based on the inspection of the $^{13}$CO(2--1) and C$^{18}$O(2--1) line data cubes, we find that the molecular gas toward G11 is well traced in a velocity range of [26, 34] km s$^{-1}$. However, the RAMPS NH$_{3}$(1, 1) emission is studied in a velocity range of [27.6, 33.7] km s$^{-1}$.  

Figures~\ref{fig5}a and~\ref{fig5}b display the $^{13}$CO and C$^{18}$O integrated intensity maps, respectively. 
A curve (in black) passing through the centers of 26 circular regions (see small filled circles in cyan) are also indicated in the $^{13}$CO 
moment-0 map (see Figure~\ref{fig4}a).
We have also produced the peak intensity maps of the $^{13}$CO and C$^{18}$O emission, which are shown in 
Figures~\ref{fig5}c and~\ref{fig5}d, respectively. 
The peak intensity map reveals the spatial distribution of the highest intensities (or peak intensities) of molecular gas emission. 
From Figures~\ref{fig5}c and~\ref{fig5}d, the areas with the highest gas concentration can be inferred toward the IRDC G11. 
The intensity-weighted velocity (moment-1) maps of the $^{13}$CO and C$^{18}$O emission are 
presented in Figures~\ref{fig5}e and~\ref{fig5}f, respectively. 
In Figures~\ref{fig5}g and~\ref{fig5}h, we display the integrated intensity map and the moment-1 map of 
the RAMPS NH$_{3}$(1, 1) emission.  

In general, the NH$_{3}$(1, 1) and C$^{18}$O emissions are known as a better dense gas tracer than $^{13}$CO. 
Hence, the entire IRDC G11 is associated with the dense gas, and contains several molecular condensations. 
In the C$^{18}$O(2--1) map, a ring-like feature (extent $\sim$130\rlap.{$''$}7 or 1.85 pc) is seen toward HFS2 (see an arrow in Figure~\ref{fig5}b). 
Note that it is not possible to identify the counterparts of the proposed IR-dark sub-filaments and HFSs in the molecular maps due to their coarse beam sizes.
In all the moment-1 maps, one can examine the gas velocity toward different parts of the `Snake' nebula with respect to almost 
its central longitude position (i.e., {\it l} = 11.14 degrees), where the molecular emission is extremely intense (see Figures~\ref{fig5}a,~\ref{fig5}b, and~\ref{fig5}g). 
All the molecular moment-1 maps clearly show a velocity variation toward the IRDC G11 (see also Section~\ref{sec0}), 
indicating the presence of two parts with different velocities (i.e., ``part-A'' and ``part-B''; see Figure~\ref{fig5}h).

\begin{figure*}
\includegraphics[width=\textwidth]{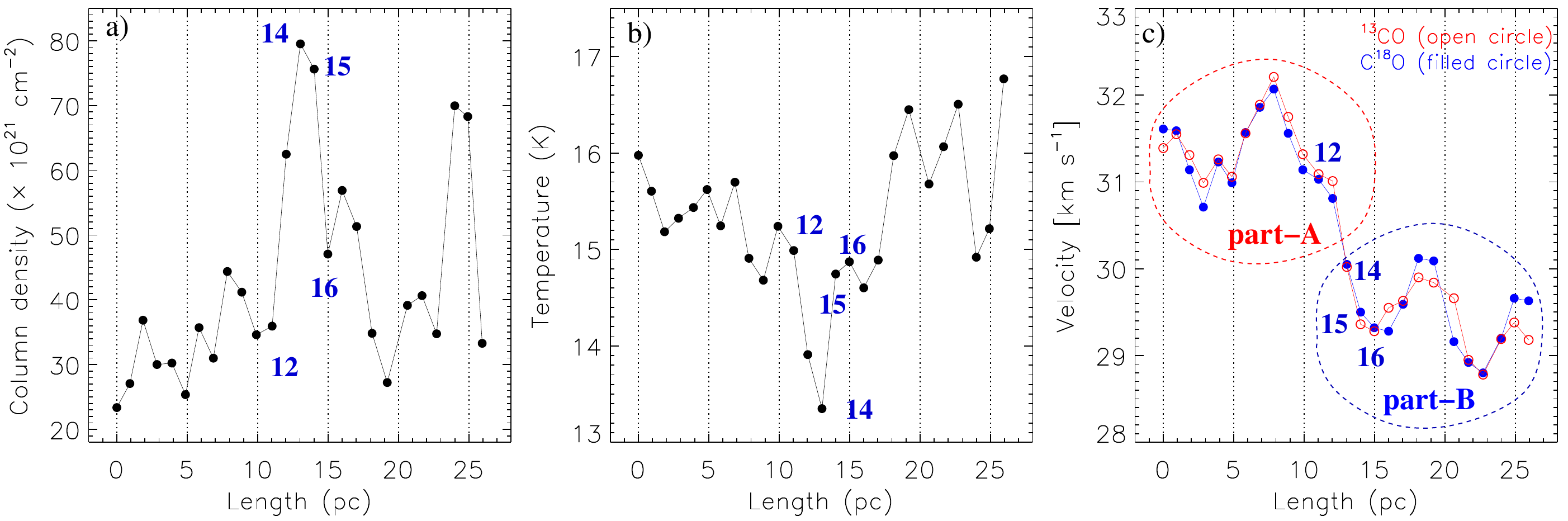}
\caption{a--c) Variation of the column density, dust temperature, and radial velocity along the IRDC G11 (see Figures~\ref{fig4}a, and~\ref{fig5}a). In the direction of 26 circular regions indicated in Figure~\ref{fig4}a, averaged dust temperatures and column
densities are determined from the {\it Herschel} temperature and column density maps, respectively, while the radial velocities are computed using the SEDIGISM $^{13}$CO and C$^{18}$O line data.} 
\label{fig6}
\end{figure*}

We examined the averaged profiles of both the $^{13}$CO and C$^{18}$O emission toward all the 26 circular regions (see Figure~\ref{fig4}a), 
allowing us to compute the peak velocity toward each circular region. 
As mentioned earlier, average values of dust temperature and column density are also computed for each circular region. 
Figures~\ref{fig6}a,~\ref{fig6}b, and~\ref{fig6}c display the variation of the column density, dust temperature, 
and radial velocity along the curve passing through the centers of 26 circular regions, respectively (see Figure~\ref{fig5}a). 
In the direction of the central part of the curve 
or the IRDC G11, high values of column density and corresponding low values of dust temperature are found. 
Using the $^{13}$CO and C$^{18}$O emission, Figure~\ref{fig6}c suggests that there may be some form of velocity oscillation toward the cloud G11. 
In the direction of the entire cloud G11, two groups of circular regions are identified with the knowledge of their peak velocities, and are labeled as ``part-A'' and ``part-B'' in Figure~\ref{fig6}c (see also Figure~\ref{fig5}h). This particular finding is also supported by the moment-1 maps of the $^{13}$CO, C$^{18}$O, and NH$_{3}$ emissions (see Figures~\ref{fig5}e,~\ref{fig5}f, and~\ref{fig5}h). 

\begin{figure}
\includegraphics[width=8.4 cm]{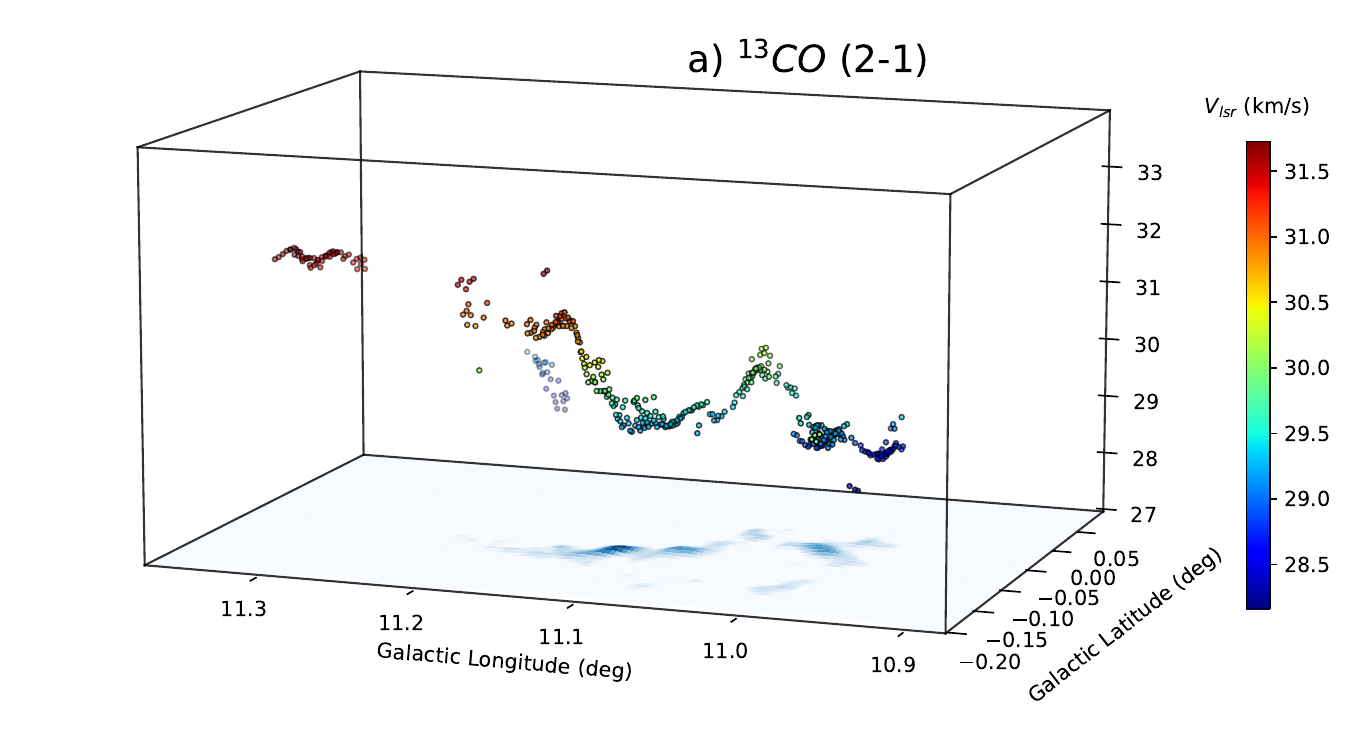}
\includegraphics[width=9.2 cm]{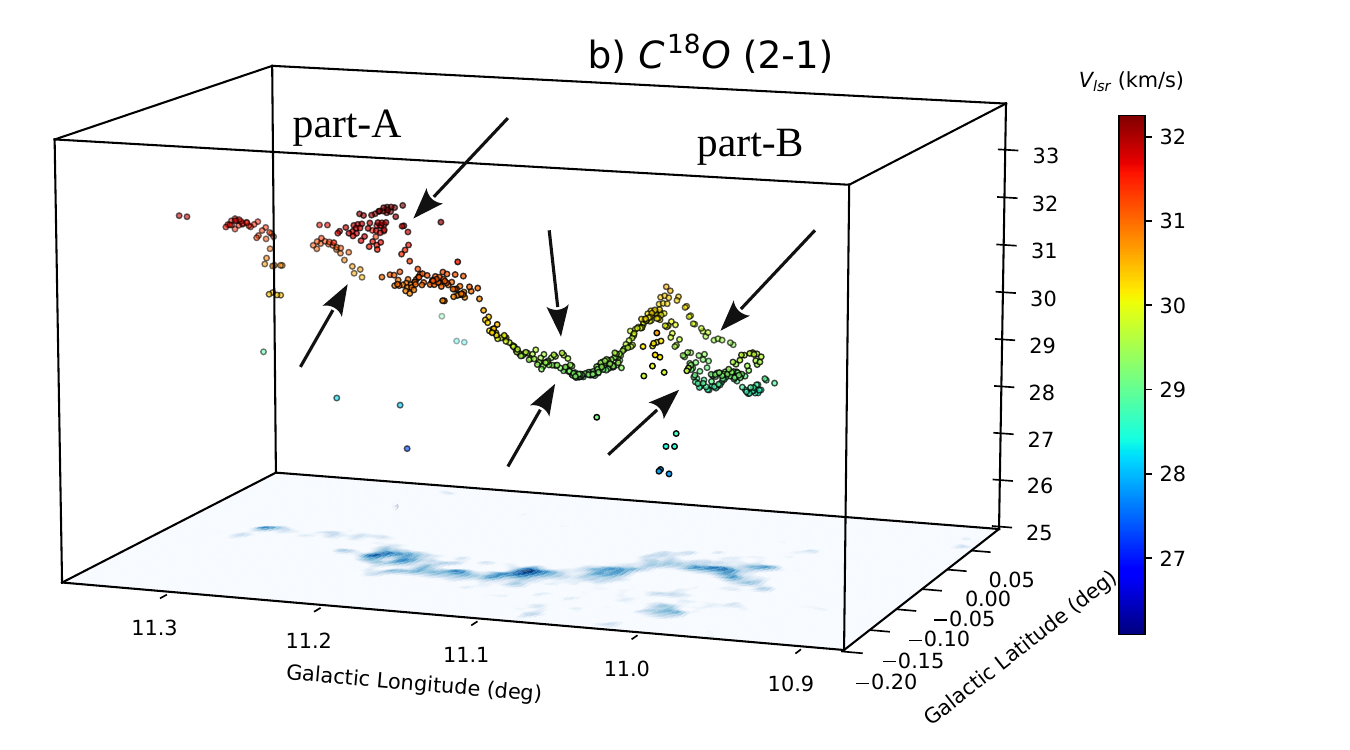}
\includegraphics[width=8.4 cm]{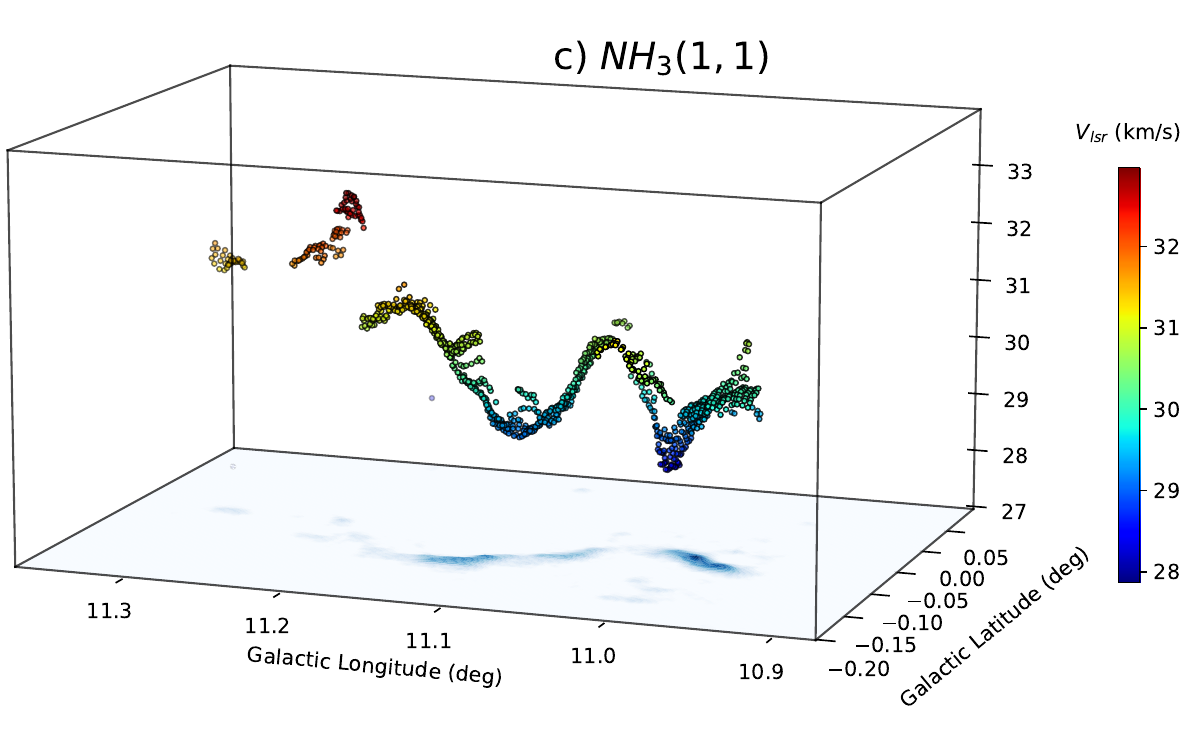}
\caption{Position-position-velocity maps of a) SEDIGISM $^{13}$CO(2--1);  b) SEDIGISM C$^{18}$O(2--1); c) RAMPS NH$_{3}$(1, 1) in the direction of G11. 
These maps are derived using the tool {\it SCOUSEPY}.}
\label{ffig8l}
\end{figure}

In Figures~\ref{ffig8l}a,~\ref{ffig8l}b, and~\ref{ffig8l}c, in order to investigate multiple velocity components toward G11, we have shown position-position-velocity (ppv) maps of the SEDIGISM $^{13}$CO(2--1), SEDIGISM C$^{18}$O(2--1), and RAMPS NH$_{3}$(1, 1) emission, respectively. These maps are generated using the tool {\it SCOUSEPY} \citep{henshaw16,henshaw19}, which is used to perform the spectral decomposition of the complex spectra. In this analysis, we define the size of Spectral Averaging Area (SAA) in pixels, over which the averaged spectra are extracted. These SAAs are regularly distributed throughout the region and cover any emission above the noise level. The pixels contained in a SAA are collectively combined to generate a spatially averaged spectrum. The averaged spectra are then automatically fitted with multiple gaussian components (if present). The peak velocity, after fitting the spectra at each SAAs, is plotted in the ppv map. Each point in the ppv map corresponds to a single SAA. We plotted the moment-0 map with emission above 3$\sigma$ in the X--Y plane at the bottom of the ppv map. This analysis is performed for the $^{13}$CO(2--1),  C$^{18}$O(2--1), and NH$_{3}$(1, 1) emission. 

Based on the inspection of the ppv maps, we find that the cloud component around 31.5 km s$^{-1}$ is dominated toward the left side of G11 (i.e., ``part-A''), while 
the right side of G11 (i.e., ``part-B'') is associated with the cloud component around 29.5 km s$^{-1}$. 
As a result, two cloud components (about 29.5 and 31.5 km s$^{-1}$) may exist, and the central area of the cloud G11 appears to be an overlapping zone of these cloud components. The presence of IR sub-filaments is also supported in the ppv maps (see arrows in Figures~\ref{ffig8l}b). 
Hence, these results enable us to suggest that the two IR sub-filaments are embedded in ``part-B'' around 29.5 km s$^{-1}$, 
while ``part-A'' around 31.5 km s$^{-1}$ also hosts the two IR sub-filaments. 
Furthermore, the ppv maps also show the noticeable velocity oscillation pattern toward G11 
as presented in Figure~\ref{fig6}c. 
We find the distribution of eight ATLASGAL dust clumps toward ``part-B'', while four clumps are present toward ``part-A''. 
Hence, there are significantly more dust clumps toward ``part-B'' compared to part-A. 
The velocity oscillation and the observed fragments/clumps appear to be related, which is detailed in Section~\ref{sec:disc}. 

\begin{figure}
\includegraphics[width=7.7cm]{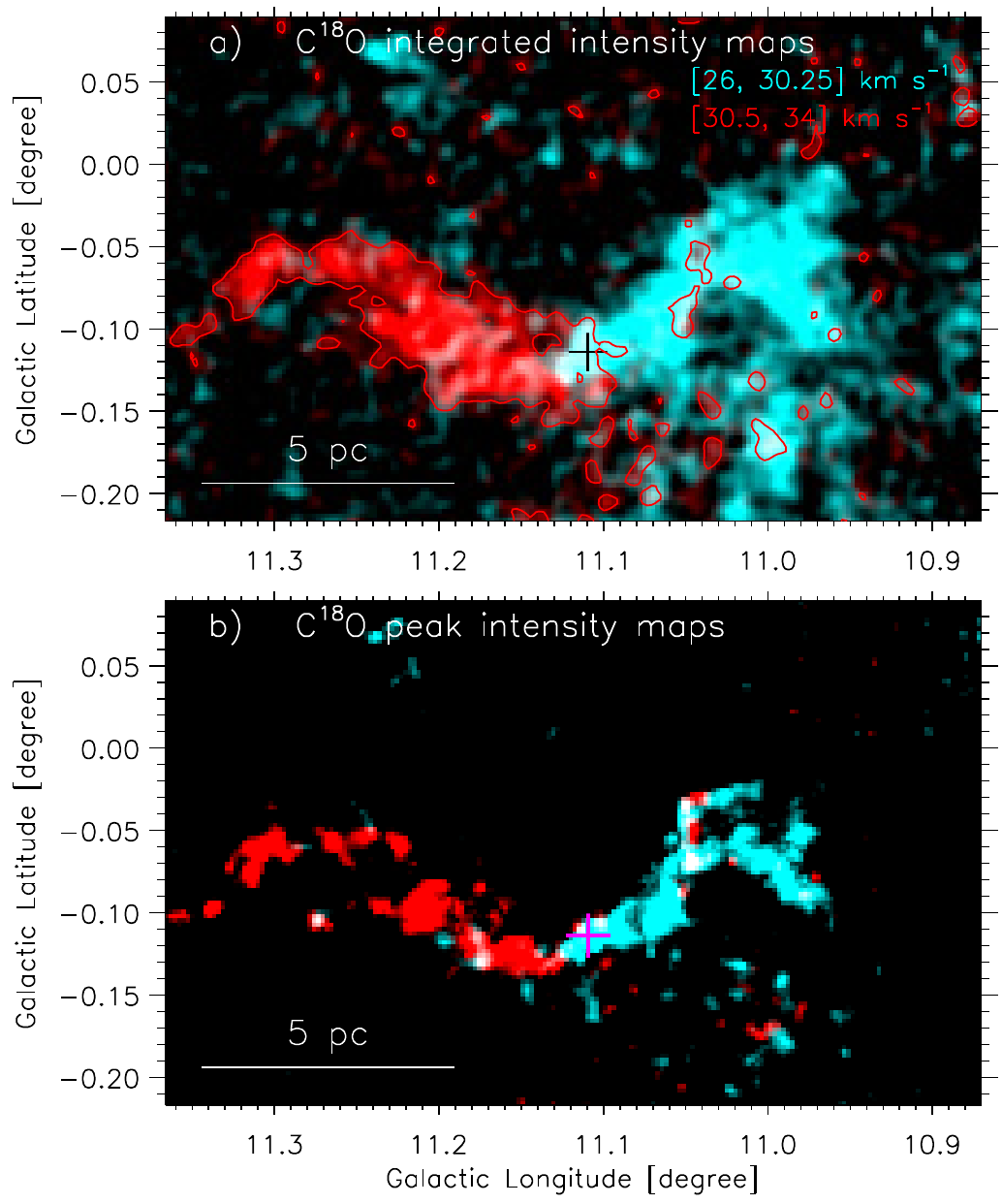}
\includegraphics[width=7.7cm]{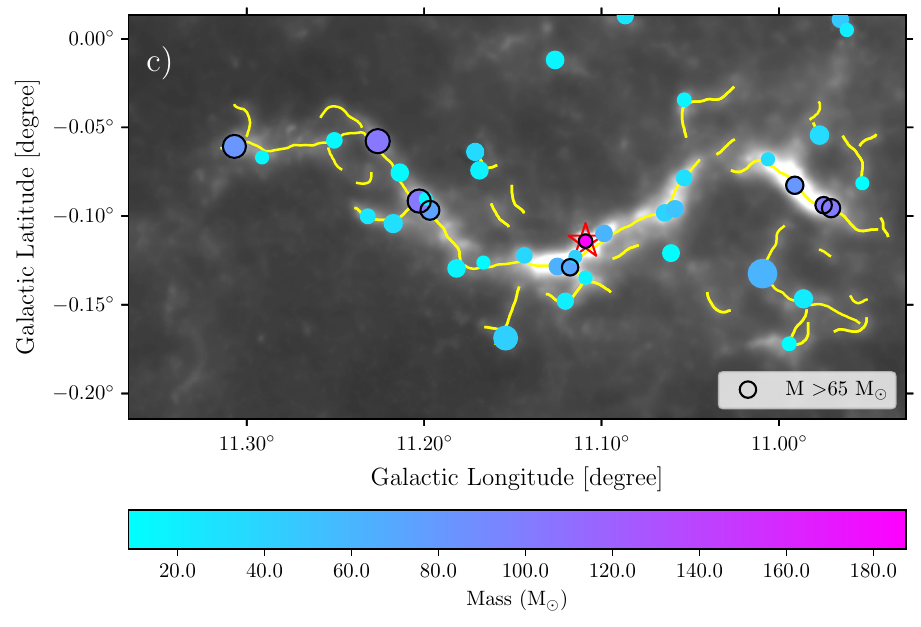}
\caption{a) Two-color composite map produced using the C$^{18}$O integrated intensity maps at [30.5, 34] and [26, 30.25] km s$^{-1}$ in red and turquoise, respectively. The red contour of the C$^{18}$O map at [30.5, 34] km s$^{-1}$ is also 
shown with a level of 1.3 K km s$^{-1}$. 
b) Same as Figure~\ref{fig8xc}a, but the composite map is made using the C$^{18}$O peak intensity maps. 
c) Overlay of the {\it getsf} extracted sources on the column density (N(H$_{2}$)) map (resolution $\sim$13\rlap.{$''$}5; see also the inset in Figure~\ref{fig4}b). 
These sources are located within the areas outlined by the contour of N(H$_{2}$) = 3.5$\times$ 10$^{22}$ cm$^{-2}$, 
and the size of dots refer to their footprint area. The encircled sources are those with a mass greater than 65 {\it M$_{\odot}$}. A source marked by a star symbol follows the MSF criteria of $M(R)> 870~M_{\odot} (R/$pc$)^{1.33}$ by \citet[][see text for more details]{Kauffmann2010}. 
Color scale refers to the mass of sources. The filament skeletons (on scales of $\sim$13\rlap.{$''$}5 to 256$''$; in yellow) 
identified using {\it getsf} are also displayed for the region where dust emission is prominent in G11. 
In panels ``a'' and ``b'', a scale bar is the same as in Figure~\ref{fig1}a.} 
\label{fig8xc}
\end{figure}

On the basis of the molecular gas distribution, two cloud components at [30.5, 34] and [26, 30.25] km s$^{-1}$ are investigated toward G11, and the spatial distribution of these cloud components using the C$^{18}$O integrated intensity emission is presented in Figure~\ref{fig8xc}a. 
In Figure~\ref{fig8xc}b, we have also shown the composite map made using the C$^{18}$O peak intensity maps at [30.5, 34] and [26, 30.25] km s$^{-1}$. 
In Figures~\ref{fig8xc}a and~\ref{fig8xc}b, most prominent overlapping zones of these cloud components are evident 
toward the central part of the IRDC G11, where the ATLASGAL clump c8, HFS3, and G11P1 are present. 

The implications of these outcomes are presented in Section~\ref{sec:disc}.  
\subsubsection{Filament skeletons and {\it getsf} continuum sources}
\label{ssubsec1a}
We have also extracted the dust continuum sources (or cores) and filament skeletons from the {\it hires} derived column density map (resolution $\sim$13$''$.5; see Figure~\ref{fig4}a). 
Figure~\ref{fig8xc}c presents the overlay of filament skeletons (scales $\sim$13$''$.5--256$''$) and {\it getsf} extracted sources on the column density map. These selected sources are distributed within the column density contour of 3.5 $\times$ 10$^{22}$ cm$^{-2}$.
One can also study the distribution of {\it getsf} continuum sources against the molecular gas from Figures~\ref{fig8xc}a and~\ref{fig8xc}b. 
To extract these structures, we used a maximum source size and filament size (i.e., FWHM, in arcsec) of 50$''$ and 100$''$, respectively. 
The filament skeletons do not clearly trace two sub-filaments in G11. Utilizing the column density map, we have estimated the mass of these sources, which varies from $\sim$9 to $\sim$188 $M_{\odot}$ with a mean (median) value of 44 (33) $M_{\odot}$ \citep[see the mass calculation in][]{dewangan23}. The physical extent (i.e., 2$R_{eff}$, where $R_{eff}=\sqrt{xy}$ is the effective radius of a source, where {\it x} and {\it y} are the semi-major axis and semi-minor axis, respectively) of these sources varies from 0.18 to 0.43 pc with a mean (median) value of 0.25 (0.24) pc.

We have further investigated the mass-radius ({\it M-R}) relationship (not shown here) of the {\it getsf} extracted sources and found 
that $M$ and $R$ are correlated by a best-fit power law with an exponent of 2.02$\pm$0.49.
This is in agreement with Larson's {\it M-R} law and previous observations of nearby clouds \citep[e.g.,][]{Lombardi2010}.
We also compared the {\it M-R} plot to the empirical {\it M-R} threshold recommended by \cite{Kauffmann2010} (i.e., $M(R)> 870M_{\odot} (R/{\rm pc})^{1.33}$), which is appropriate for dense sources prone to forming massive stars. We found that only one source follows the {\it M-R} criteria of MSF (i.e., G11P1; see a star in Figure~\ref{fig8xc}c) . However, we note that most massive sources lie toward the main spine of G11, which is evident from the 
distribution of sources (mass $>$ 65 {\it M$_{\odot}$}) and filament skeletons in Figure~\ref{fig8xc}c. 
\subsubsection{Estimation of molecular cloud mass}
\label{sssubsec2}
We have calculated the gas mass of the elongated structure traced in the molecular maps (see an area indicated by a dashed curve in Figure~\ref{fig5}f). 
The $^{13}$CO(2--1) emission is well distributed within this selected area, and the dense gas tracer C$^{18}$O emission is also found within the area. 
The optical depth ($\tau$) for the $^{13}$CO(2--1) emission has been calculated using the equation \citep{Liu_2020_apj} 
 \begin{equation}
    \frac{T_r{\rm (^{13}CO)}}{T_r{\rm (C^{18}O)}}\approx\frac{1-{\rm exp} (-\tau_{13})}{1-{\rm exp} (-\tau_{13}/R_{iso})}, \label{eq-tau}
 \end{equation}
where $T_r$ is the measured brightness temperature and $R_{iso}$ is the isotope ratio between $^{13}$CO and C$^{18}$O. $R_{iso}$ is assumed to be 7.4 \citep{Areal_2018}.\\ 

Assuming the system in local thermodynamic equilibrium (LTE), we can calculate the column density of a linear molecule using the equation \citep{Mangum_2015_pasp}  
 \begin{eqnarray}
    N&=&\frac{3h}{8\pi^3\mu^2S}\frac{Q_\mathrm{rot}}{g_{J}}
    \frac{\mathrm{exp}\left(\frac{E_\mathrm{up}}{k T_\mathrm{ex}}\right)}{\mathrm{exp}\left(\frac{h\nu}{k T_\mathrm{ex}}\right) - 1} \nonumber \\
    & &\times\frac{1}{J(T_\mathrm{ex})-J(T_\mathrm{bg})}\frac{\tau}{1-\mathrm{exp}(-\tau)}\int T_\mathrm{r}~dv \label{eq-lte-density}.
 \end{eqnarray}
Here $J(T) = \frac{h\nu/k}{\mathrm{exp}(h\nu/k T)-1}$ and $T_{\rm bg}$ = 2.73~K is the cosmic microwave background temperature. The excitation temperature ($T_\mathrm{ex}$) for the target source is assumed to be 10\thinspace K. The symbol $\mu$ stands for the dipole moment of the molecule. The degeneracy ($g_\mathrm{J})$ is $2J_u+1 = 5$, and the line strength ($S$) is $\frac{J_u}{2J_u+1} = \frac{2}{5}$, where $J_u$ is the rotational quantum number of the upper state (i.e., $J_u = 2$ for the $J = 2-1$ transition). $E_\mathrm{up}$ is the energy of the upper state. We have adopted an approximated formula for the rotational partition function ($Q_\mathrm{rot}$) mentioned in various previous studies \citep[e.g.,][]{McDowell_1988,Mangum_2015_pasp,Yuan_2016_apj}. Other molecular parameters have been adopted from the Jet Propulsion Laboratory (JPL)
Molecular Spectroscopy database and spectral line catalog \citep{Pickett_1998}. 

With the help of our dervied $\tau_{13}$ map (see Equation~\ref{eq-tau}) and the integrated $^{13}$CO(2--1) emission at [26, 34]~km s$^{-1}$ in Equation~\ref{eq-lte-density}, the $^{13}$CO column density map is produced. Now considering the column density ratio between H$_2$ and $^{13}$CO \citep[i.e., $\frac{N(\mathrm{H_2})}{N(\mathrm{^{13}CO})}$ = $7 \times 10^5$;][]{Frerking_1982apj}, we have obtained the $N(\mathrm{H_2})$ map. Considering the distance (= 2.92~kpc) and the mean molecular weight \citep[= 2.8;][]{Kauffmann_2008}, we have determined the total mass of the elongated cloud (see a dashed curve in Figure~\ref{fig5}f) 
to be $\sim$10.6~$\times 10^4$($\sim$9~$\times 10^4$) {\it M$_{\odot}$} at $T_\mathrm{ex}$ = 10(20)~K. 
From Figure~\ref{fig4}b, an average value of the dust temperature of the `Snake' nebula is estimated to be $\sim$16~K. 
Using this value ($T_\mathrm{ex}$ = 16~K), we have also computed the total mass of the elongated 
cloud to be $\sim$8.9~$\times 10^4$ {\it M$_{\odot}$}.  

We have also used the C$^{18}$O(2--1) line data for determining the column density map and the cloud mass. 
In the case of the optically thin C$^{18}$O(2--1) line, the term $\frac{\tau}{1-\mathrm{exp}(-\tau)}$ tends to be 1 (see Equation~\ref{eq-lte-density}). 
Now similar to the $^{13}$CO analysis, the integrated C$^{18}$O(2--1) emission at [26, 34]~km s$^{-1}$ in Equation~\ref{eq-lte-density} enables us to produce the C$^{18}$O column density map. From the column density ratio between H$_2$ and C$^{18}$O \citep[i.e., $\frac{N(\mathrm{H_2})}{N(\mathrm{C^{18}O})}$ = $5.8\times 10^6$;][]{Frerking_1982apj}, we have obtained the $N(\mathrm{H_2})$ map. 
Using the C$^{18}$O(2--1) line data, the mass of the elongated structure (see a dashed curve in Figure~\ref{fig5}f) is computed 
to be $\sim$4.3~$\times 10^4$($\sim$3.6~$\times 10^4$ and $\sim$3.7~$\times 10^4$) {\it M$_{\odot}$} at $T_\mathrm{ex}$ = 10(16 and 20)~K. 
Using the {\it Herschel} column density map, the cloud mass is estimated to be $\sim$5.5 $\times$ 10$^{4}$ {\it M$_{\odot}$} (see Section~\ref{subsec:temp}). 
Note that the C$^{18}$O emission is known to trace relatively higher density regions in comparison to the $^{13}$CO emission (and/or dust column density map). 
Consequently, the mass derived from C$^{18}$O is expected to be lower than the estimated mass from $^{13}$CO. It is clearly reflected in our results.

Apart from the mass of the entire cloud, using the C$^{18}$O emission, we have also computed the masses of two cloud components at [30.5, 34] and [26, 30.25] km~s$^{-1}$ as presented in Figure~\ref{fig8xc}a.
The masses of the clouds at [30.5, 34] km s$^{-1}$ and [26, 30.25] km s$^{-1}$ are estimated 
to be $\sim$1.9~$\times 10^4$($\sim$1.62~$\times 10^4$ and $\sim$1.64~$\times 10^4$) and 
$\sim$2.4~$\times 10^4$($\sim$2.0~$\times 10^4$ and $\sim$2.0~$\times 10^4$) {\it M$_{\odot}$} at $T_\mathrm{ex}$ = 10(16 and 20)~K, respectively. The primary sources of uncertainty in the calculations of mass are involved in $\frac{N(\mathrm{H_2})}{N(\mathrm{^{13}CO})}$ or $\frac{N(\mathrm{H_2})}{N(\mathrm{C^{18}O})}$, the estimation of the distance, and the observational random errors. These factors can result in an uncertainty of anywhere from 30\% to 50\% \citep[e.g.,][]{bhadari22}. 
\subsection{Inner environment of G11P1 and G11P6}
\label{xbsec3}
\subsubsection{\emph{JWST} NIR view of G11P1 and G11P6}
\label{nircamsc}
We have examined the \emph{JWST} NIRCam images, which are available only toward G11P1/c7 and G11P6/HFS4/c2 (see the dot-dashed box in Figures~\ref{fig3}b and~\ref{fig3}d).
In Figure~\ref{fig3lc}a, in the direction of G11P1, we display a three-color composite map made using the \emph{JWST} F444W (in red), F356W (in green), and F200W (in blue) images. The composite map reveals a central region, which is surrounded by at least five small scale filaments (length $<$ 0.35 pc) in absorption . 
Such embedded configuration around G11P1 is also evident in the ratio map of F444W 
($\lambda_{eff}$/$\Delta \lambda$: 4.421/1.024 $\mu$m) and F356W ($\lambda_{eff}$/$\Delta \lambda$: 3.563/0.787 $\mu$m) images (see Figure~\ref{fig3lc}b). 
The central region contains some point-like sources that are saturated in the images. 
Additionally, we find the presence of several embedded sources that are detected only in the \emph{JWST} F444W image and are seen toward the small scale filaments. 
From Figure~\ref{fig3lc}a and~\ref{fig3lc}b together, we have discovered a small scale IR-dark HFS candidate (extent $\sim$0.55 pc) toward G11P1 (i.e., G11P1-HFS; see also Figures~\ref{zzfig10}c--\ref{zzfig10}e). 

Figure~\ref{fig3lc}c displays a three-color composite map produced using the \emph{JWST} F444W (in red), F356W (in green), and F200W (in blue) images toward G11P6 or HFS4. 
The map reveals a small scale IR-dark HFS candidate (extent $<$ 0.8 pc) around 
a bright and saturated source that is classified as a Class~I protostar (see Section~\ref{subsec:temp}). 
The composite map is also overlaid with the positions of the dust condensations (or continuum sources; see hexagons in Figure~\ref{fig3lc}c) traced in the Submillimeter Array (SMA) 880 $\mu$m continuum map \citep[see Table~3 in][]{wang14}. \citet{wang14} reported that some of these dust continuum sources are associated with star formation activities. Hence, it is likely that the SMA continuum sources are distributed toward the central hub of this small scale IR-dark HFS candidate. 
Using the \emph{JWST} F200W image, a zoomed-in view of 
the Class~I protostar is presented in Figure~\ref{fig3lc}d. 
The dust cocoon or envelope-like feature (extent $<$ 0.15 pc) is investigated around the Class~I protostar. 
In Figure~\ref{fig3lc}d, we have also highlighted small scale filaments (in absorption) visually seen in the \emph{JWST} images (see dashed lines). 
In the north-east of the bright source, dark regions and embedded nebulus features traced in the \emph{JWST} F444W image are observed. 
The \emph{JWST} images also favour the presence of HFS4 toward G11P6 as seen in the {\it Spitzer} images.
\subsubsection{ALMA 1.16 mm continuum map}
\label{consec2}
Figure~\ref{fig10}a displays the ALMA 1.16 mm continuum map and the 1.16 mm continuum contours.
A dusty envelope-like feature in the inner 18000 AU is evident in the map, 
and is surrounded by several finger-like features (extent $\sim$3500--10000 AU; see arrows in Figure~\ref{fig10}a). 
In the continuum map, we find a few continuum sources/peaks located within the dusty envelope-like feature. 
Hence, hierarchical structures in G11P1 are evident in the ALMA continuum map (see Figure~\ref{fig10}b). Furthermore, we have also identified a small-scale filamentary-like feature (ssff; extent $\sim$0.15 pc), which appears to be connected with the envelope-like feature. 
A diffuse feature labeled as ``vssff'' is also indicated in Figure~\ref{fig10}a. The envelope-like feature is seen between ``ssff'' and ``vssff''. 
Both these features ``ssff'' and ``vssff'' seem to be connected with the IR filaments in the large-scale view ($>$ 1 pc).    

The {\it clumpfind} IDL program \citep{williams94} has been employed in the ALMA continuum map at 1.16 mm to depict continuum sources. This particular analysis gives the total flux, the FWHM not corrected for beam size for the x-axis (i.e., FWHM$_{x}$), and for the y-axis (i.e., FWHM$_{y}$) of each identified source. Figure~\ref{fig10}b shows clumpfind decomposition of the ALMA 1.16 mm continuum emission, where spatial boundaries of continuum sources can be examined. We have shown a contour (in navy blue) at [0.03] $\times$ 10.4 mJy beam$^{-1}$ 
and a contour (in pink) at [0.055] $\times$ 10.4 mJy beam$^{-1}$ in Figure~\ref{fig10}b. The entire extended structure containing the envelope-like feature and the ``ssff'' is traced by the contour at [0.03] $\times$ 10.4 mJy beam$^{-1}$, which is labeled as the source ``A''. 
Using the contour (in pink) at [0.055] $\times$ 10.4 mJy beam$^{-1}$, we have identified three continuum sources ``B--D'' in the map. 
Three continuum sources ``b1--b3'' are investigated toward the continuum source ``B''. 
Table~\ref{tab3} provides the fluxes, deconvolved FWHM$_{x}$ \& FWHM$_{y}$, and masses of all the continuum sources marked in Figure~\ref{fig10}b. 

The following expression is utilized to compute the mass of each continuum source \citep{hildebrand83}: 
\begin{equation}
M \, = \, \frac{D^2 \, F_\nu \, R_t}{B_\nu(T_d) \, \kappa_\nu}
\end{equation} 
\noindent where $F_\nu$ is the total integrated flux (in Jy), 
$D$ is the distance (in kpc), $R_t$ is the gas-to-dust mass ratio, 
$B_\nu$ is the Planck function for a dust temperature $T_d$, 
and $\kappa_\nu$ is the dust absorption coefficient. 
In this work, we adopted $\kappa_\nu$ = 1.13\,cm$^2$\,g$^{-1}$ at 1.1 mm \citep{ossenkopf94,contreras18}, 
$T_d$ = [10, 15, 25]~K, and $D$ = 2.92 kpc. Note that previously, \citet{wang14} considered the average temperature of the clump hosting G11P1 to be 15~K. The mass of the continuum source ``A'' containing the envelope-like feature and the ``ssff'' 
is estimated to be [95.5, 50, 25] {\it M$_{\odot}$} at $T_d$ = [10, 15, 25]~K. 
We have estimated masses of three continuum sources (i.e., b1, b2, and b3) located 
within the dusty envelope-like feature, which are [14.6, 7.6, 3.8], [28.3, 14.8, 7.4], and [7.8, 4.1, 2.0] {\it M$_{\odot}$} 
at $T_d$ = [10, 15, 25]~K, respectively (see Table~\ref{tab3}). 

Independently, we have also employed the python-based {\it astrodendro} tool \citep[e.g.,][]{Rosolowsky2008} to 
identify sub-structures in the ALMA 1.16 mm continuum map \citep[see also][]{bhadari23}. 
The input parameters, namely ``min\_value'' and ``min\_delta'', 
were set to 0.5 and 0.2 mJy beam$^{-1}$, respectively. 
Additionally, the parameter ``min\_npix'' was chosen to ensure that the dendrogram structures contain at least 2 ALMA beams.
The resulting dendrogram consists of 4 leaf and 3 branch structures (size $\sim$0.01 to 0.1 pc). 
In Figure~\ref{fig10}c, we have marked four leaves and one branch (see ellipses).
The branch is indicated by the structure L1 (see the dot-dashed ellipse), while leaves are represented by structures I1--I4 (see solid ellipses). 
The total fluxes of L1, I1, I2, I3, and I4 are 154.55, 18.32, 13.34, 34.62, and 8.83 mJy, respectively. 
The dendrogram structures L1, I3, and I4 have fluxes that are similar to the {\it clumpfind} 
sources B, C, and D, respectively (see Figures~\ref{fig10}b and~\ref{fig10}c). 
Dendrogram leaves I1 (size $\sim$0\rlap.{$''$}83 $\times$ 0\rlap.{$''$}52)  and I2 (size $\sim$0\rlap.{$''$}96 $\times$ 0\rlap.{$''$}45) appear spatially more compact toward the {\it clumpfind} sources b1 and b2, respectively. 
The masses of I1 and I2 are computed to be [6.4, 3.3, 1.7] and [4.6, 2.4, 1.2] {\it M$_{\odot}$} at $T_d$ = [10, 15, 25]~K, 
respectively. 
It is worth noting that the results obtained from both the {\it clumpfind} algorithm 
and the {\it astrodendro} tool exhibit close agreement with each other. 
One may consider the uncertainty in the estimated mass to be typically $\sim$20\% to $\sim$50\%. 
This expected error in the mass calculation may be due to uncertainties in the opacity, dust temperature, and measured flux. 
\subsubsection{ALMA H$^{13}$CO$^{+}$(3--2) line data}
\label{linsec2}
\begin{figure*}
\includegraphics[width=\textwidth]{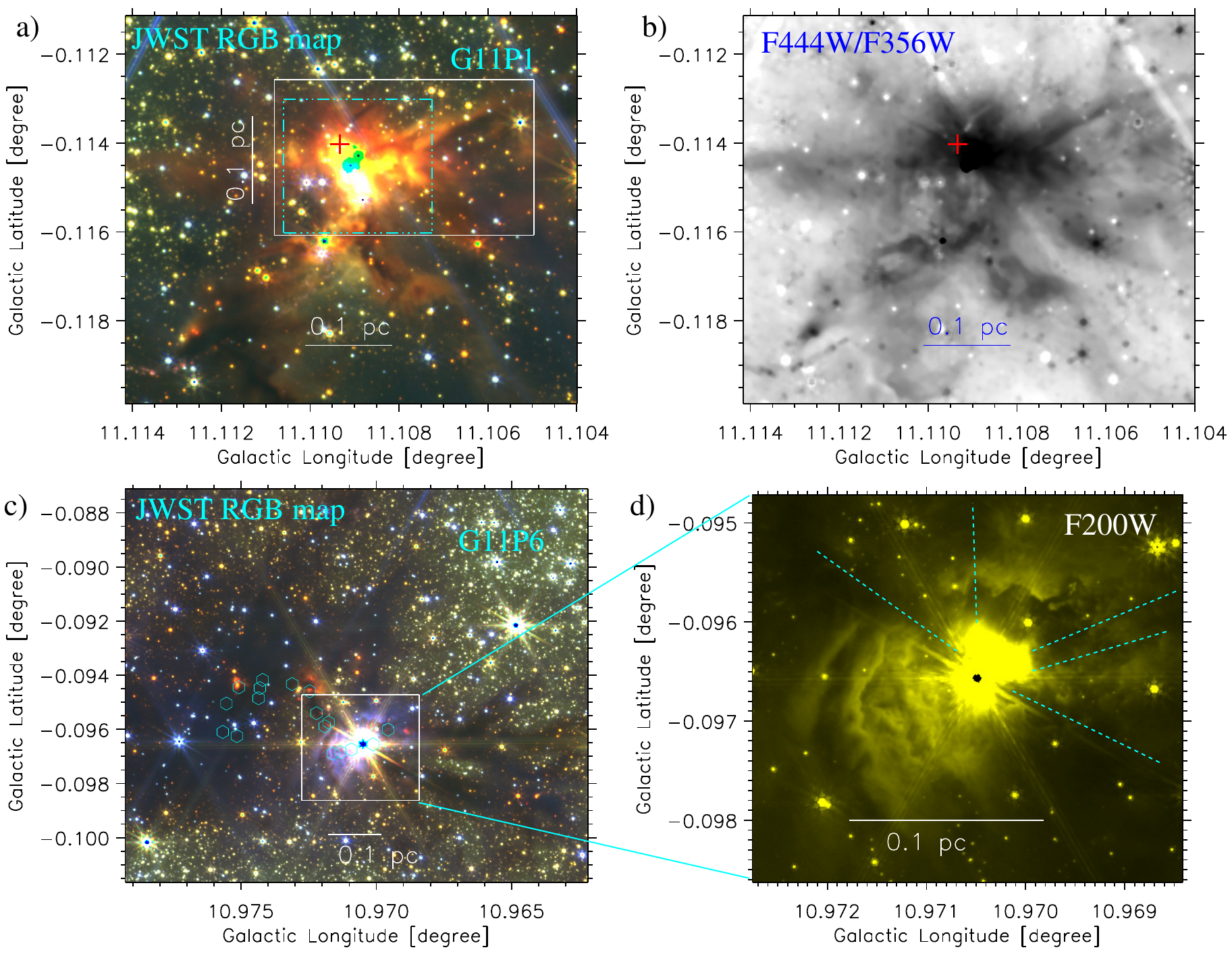}
\caption{a) \emph{JWST} three-color composite map (F444W (in red), F356W (in green), and F200W (in blue) images in square root scale).  
The presented area is selected in the direction of the clump~c7 or G11P1 (see Figure~\ref{fig3}d). 
The solid box is presented as a zoomed-in view in Figure~\ref{fig10}, while the dot-dashed box is shown as a zoomed-in view in Figure~\ref{zzfig10}. 
b) \emph{JWST} F444W/F356W (in linear scale). 
The \emph{JWST} ratio image is exposed to median filtering with a width of 6 pixels and smoothing by 3 $\times$ 3 pixels using the ``box-car'' algorithm. 
c) Overaly of the positions of the SMA 880 $\mu$m continuum sources \citep[see hexagons; from][]{wang14} on the \emph{JWST} three-color composite map (F444W (in red), F356W (in green), and F200W (in blue) images in square root scale). 
The presented area is selected toward the clump~c2 or G11P6 (see Figure~\ref{fig3}b).  
d) \emph{JWST} F200W (in square root scale; see the solid box in Figure~\ref{fig3lc}c). It is a zoomed-in view of an area toward the clump c2. 
Dashed lines highlight embedded small scale filaments. In panels ``a'' and ``b'', a cross highlights the position of the 6.7 GHz methanol maser. In each panel, a scale bar represents a spatial scale at a distance of 2.92 kpc.}
\label{fig3lc}
\end{figure*}

\begin{figure*}
\includegraphics[width=15.2 cm]{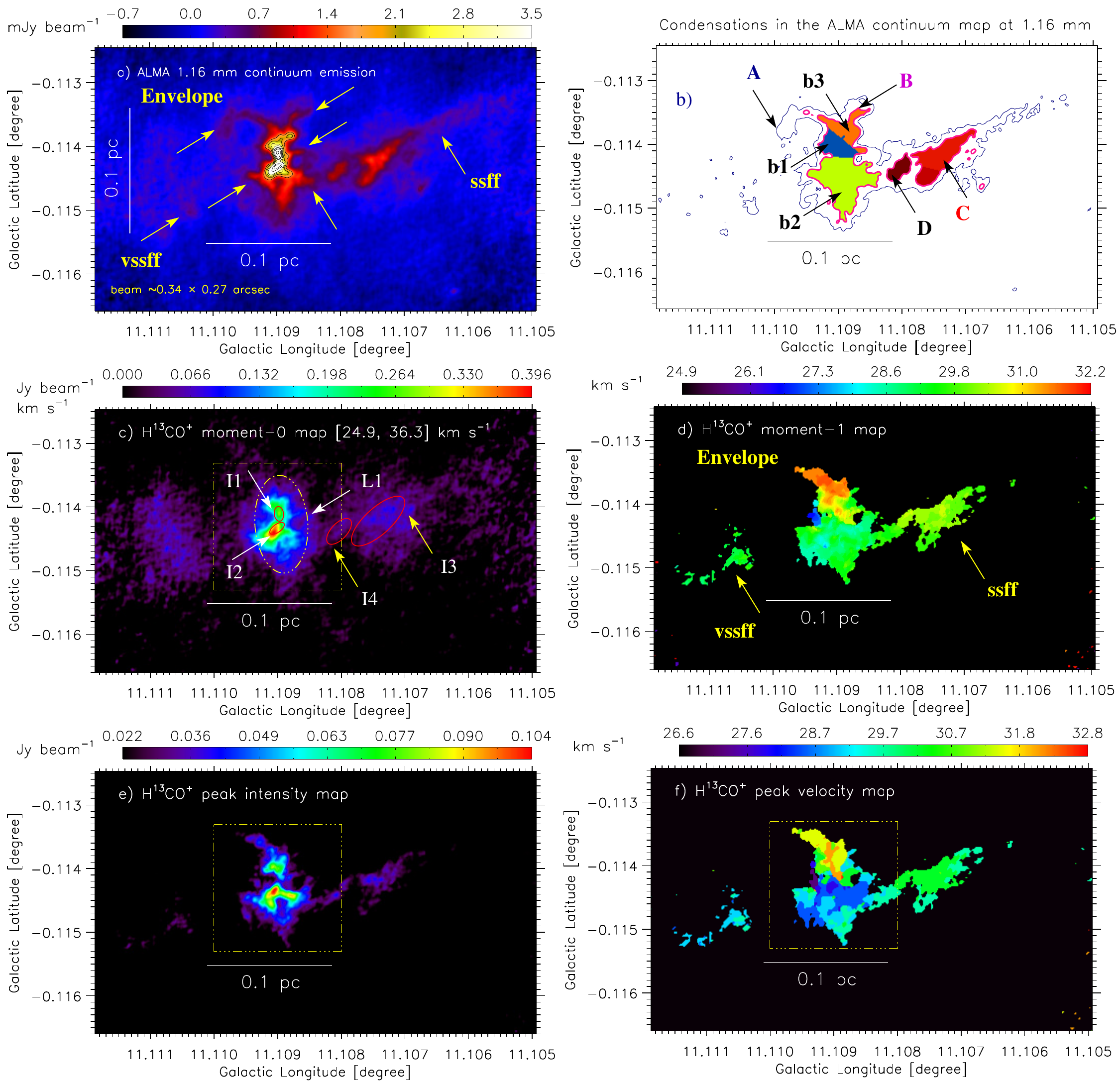}
\includegraphics[width=10 cm]{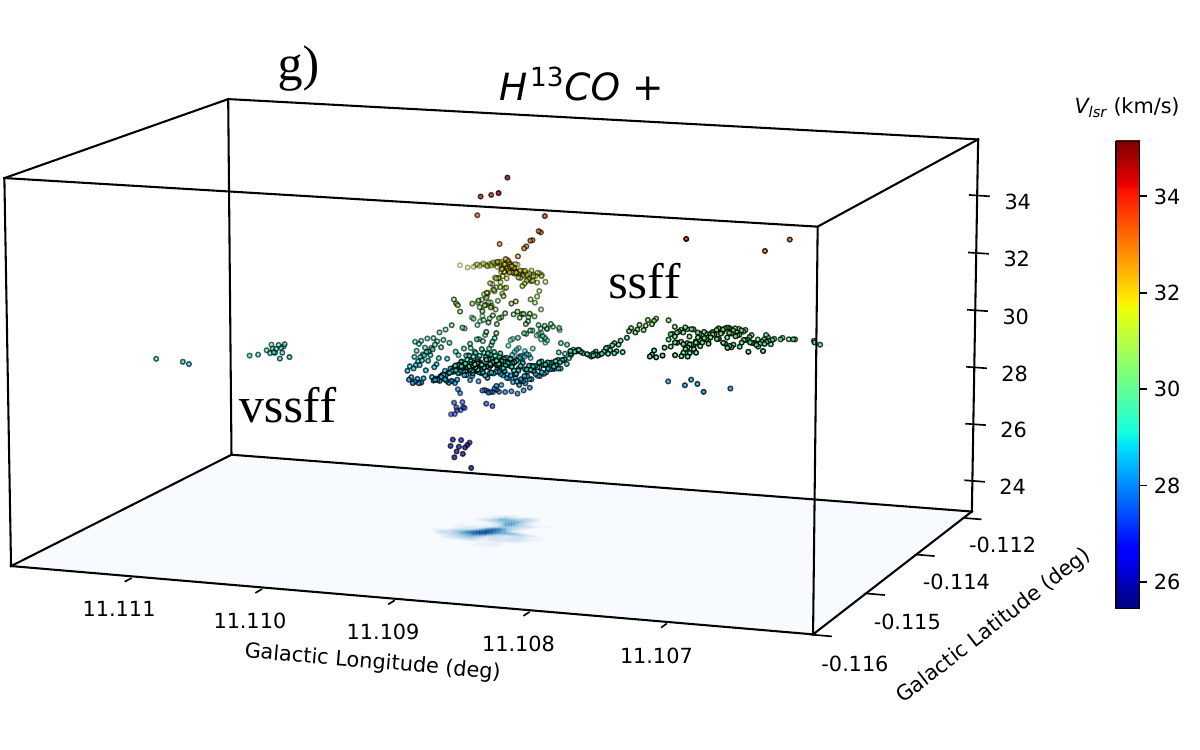}
\caption{a) ALMA 1.16 mm continuum map of G11P1 (see the solid box in Figure~\ref{fig3}d).  
The map is also overlaid with the 1.16 mm continuum contours, and the contour levels are (0.15, 0.2, 0.25, 0.3, 0.4, 0.45, 0.6, 0.9) $\times$ 
10.4 mJy beam$^{-1}$ (where 1$\sigma$ $\sim$185 $\mu$Jy beam$^{-1}$). 
b) Clumpfind decomposition of the ALMA 1.16 mm continuum emission, allowing us to infer the spatial 
boundaries of selected continuum sources (see labels and also Table~\ref{tab3}). 
The navy blue contour at [0.03] $\times$ 10.4 mJy beam$^{-1}$ and the pink contour at [0.055] $\times$ 10.4 mJy beam$^{-1}$ are also presented. 
c) ALMA H$^{13}$CO$^{+}$(3--2) moment-0 map at [24.9, 36.3] km s$^{-1}$. The locations of Dendrogram structures (i.e., leaves and branches) using the ALMA 1.16 mm continuum map are also marked. The object L1 represents one branch (see the dot-dashed ellipse), while the objects I1--I4 show leaves (see solid ellipses). 
d) ALMA H$^{13}$CO$^{+}$(3--2) moment-1 map. 
e) ALMA H$^{13}$CO$^{+}$(3--2) peak intensity map. 
f) ALMA H$^{13}$CO$^{+}$(3--2) peak velocity map. 
g) PPV map of the ALMA H$^{13}$CO$^{+}$(3--2) emission. The map is produced using the tool {\it SCOUSEPY}. 
In panels ``c'', ``e'', and ``f'', the dot-dashed box emcopasses the area presented in Figures~\ref{fig11}a--\ref{fig11}f. 
In panels ``a--f'', the scale bar shows a spatial scale of 0.1 pc at a distance of 2.92 kpc.} 
\label{fig10}
\end{figure*}

In Figure~\ref{fig10}c, we show the ALMA H$^{13}$CO$^{+}$(3--2) moment-0 map at [24.9, 36.3] km s$^{-1}$, tracing the dense molecular gas toward the envelope-like feature, ``ssff'', and ``vssff''.  
The IR emission is also depicted toward ``ssff'' (see Figure~\ref{fig3lc}b). Using the H$^{13}$CO$^{+}$ emission, the velocity field toward the envelope-like feature and the ``ssff'' can be studied in the moment-1 map (see Figure~\ref{fig10}d). A noticeable velocity variation is clearly evident in the moment-1 map. We have also generated the peak intensity map and the peak velocity (or velocity at peak intensity) map of the H$^{13}$CO$^{+}$(3--2) emission, 
which are presented in Figures~\ref{fig10}e and~\ref{fig10}f, respectively. 
Both the ALMA continuum map and the H$^{13}$CO$^{+}$(3--2) line data reveal almost similar morphologies in the inner environment ($<$ 20000 AU) of G11P1. 
In other words, the H$^{13}$CO$^{+}$(3--2) line data also support the existence of a configuration that includes the envelope-like feature that is observed between ``ssff'' and ``vssff''.  

In Figure~\ref{fig10}e, the locations, where the gas is most concentrated, are revealed. 
The peak velocity map enables us to obtain the gas velocity distribution at the locations of peak intensity. 
Figure~\ref{fig10}f displays the dominant velocities of the gas, providing insights into gas kinematics and motion. 
In Figure~\ref{fig10}f, one can examine the gas velocity toward different parts of the envelope-like feature with respect to its central part, 
where the intensity of the H$^{13}$CO$^{+}$ emission is very high (see Figures~\ref{fig10}c and~\ref{fig10}e). 
A significant velocity variation toward the envelope-like feature is evident in the peak velocity map (see also moment-1 map in Figure~\ref{fig10}d). 
This argument is also supported by the ppv map of the H$^{13}$CO$^{+}$ emission (see Figure~\ref{fig10}g). 
In Figure~\ref{fig10}g, the emission at the bottom (or X--Y plane) is the moment-0 map. 
In the direction of the feature ``ssff'', the sub-structures are evident in both physical and velocity space. 
For the spectral decomposition of the H$^{13}$CO$^{+}$ spectra, we used the SAAs of size 3 $\times$ 3 
pixels in the tool {\it SCOUSEPY}.  

\begin{figure*}
\includegraphics[width=\textwidth]{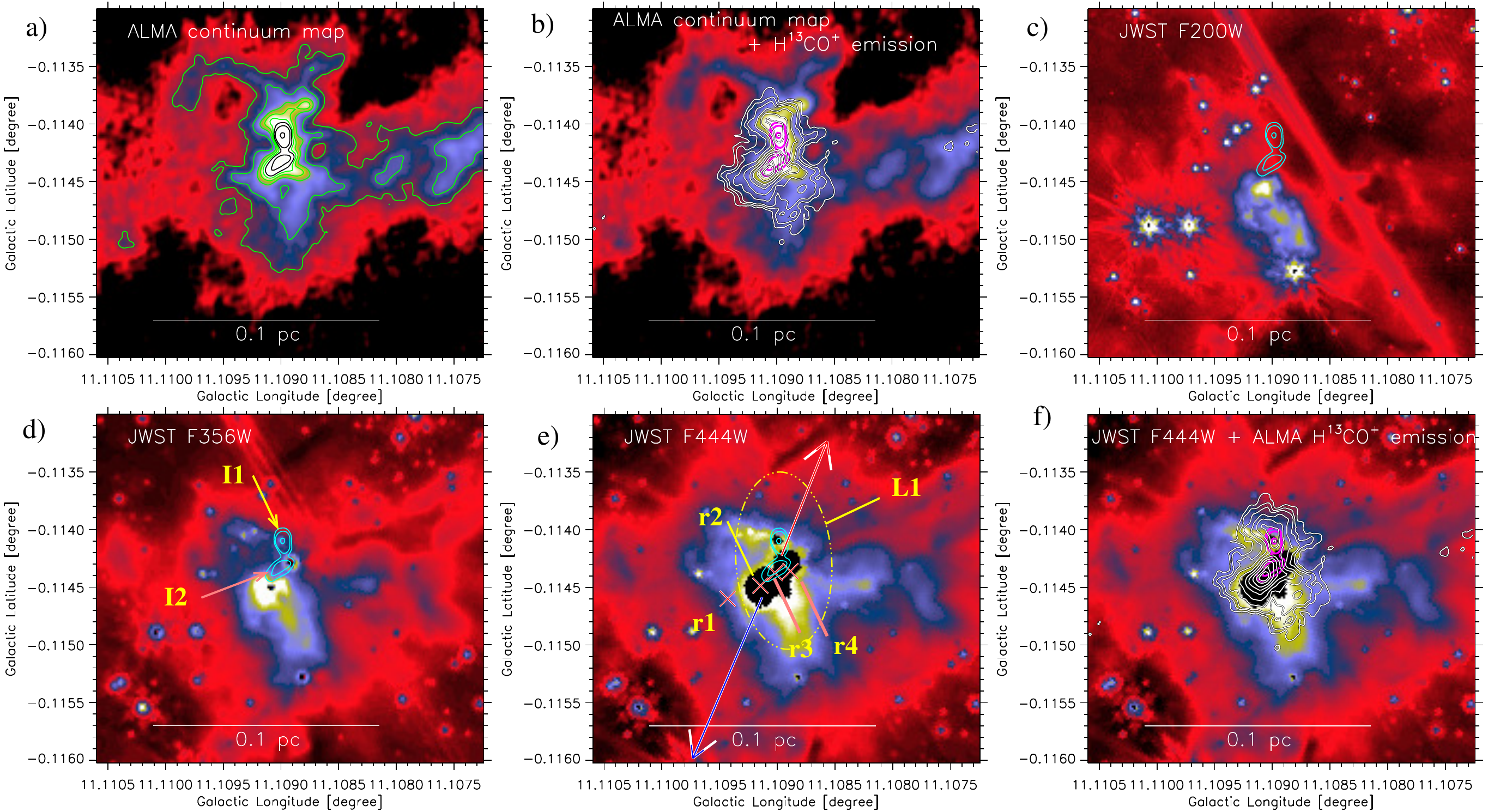}
\caption{A zoomed-in view of G11P1 using the {\it JWST}  and ALMA images (see a dot-dashed box in Figure~\ref{fig3lc}a). 
a) ALMA 1.16 mm continuum map overlaid with the 1.16 mm continuum emission 
contours (in green and black). The levels of the contours (in green) 
are (0.038, 0.08, 0.15, 0.2, 0.25) $\times$ 10.2 mJy beam$^{-1}$. b) ALMA 1.16 mm continuum map overlaid with the ALMA 1.16 mm continuum 
emission contours (in magenta) and the ALMA H$^{13}$CO$^{+}$(3--2) emission contours (in white). 
The levels of  the ALMA H$^{13}$CO$^{+}$(3--2) emission contours
 (in white) are (0.2, 0.25, 0.3, 0.4, 0.5, 0.6, 0.7, 0.8, 0.9, 0.95)  $\times$ 395 mJy beam$^{-1}$ km s$^{-1}$.
Overlay of the ALMA 1.16 mm continuum emission contours (in cyan) on c) {\it JWST} F200W image; 
d) {\it JWST} F356W image; e) {\it JWST} F444W image. 
f) {\it JWST} F444W image overlaid with the ALMA 1.16 mm continuum emission contours (in magenta) and the ALMA H$^{13}$CO$^{+}$(3--2) emission contours (in white).
In all panels, the levels of the ALMA continuum emission contours (in black, magenta, cyan) are 
(0.3, 0.4, 0.9) $\times$ 10.2 mJy beam$^{-1}$, and indicate the locations of two objects (I1 and I2) as marked in Figure~\ref{fig10}c. 
In panel ``e'', the object L1 highlighted by a dot-dashed ellipse is the same as indicated in Figure~\ref{fig10}c. 
Four radio continuum sources at 4.9 GHz \citep[see Table~1 in][]{rosero14} are indicated by multiplication symbols, and are labeled as r1--r4 in 
panel ``e''. In panel ``e'', arrows highlight the direction of a previously reported SiO(5--4) outflow \citep[see Figure~3 in][]{wang14}. 
In each panel, the scale bar shows a spatial scale of 0.1 pc at a distance of 2.92 kpc.}  
\label{zzfig10}
\end{figure*}

We have also examined the {\it JWST} NIR images and the ALMA line and continuum maps together 
toward G11P1 in Figure~\ref{zzfig10}. The envelope-like feature traced in the ALMA continuum map is presented in Figure~\ref{zzfig10}a, 
and the distribution of the ALMA H$^{13}$CO$^{+}$(3--2) emission toward this feature 
is displayed in Figure~\ref{zzfig10}b. 
From Figure~\ref{zzfig10}b, it is found that the peaks of the continuum emission (see magenta contours) or ALMA continuum sources I1 and I2 coincide with the ALMA H$^{13}$CO$^{+}$(3--2) emission peaks (see white contours). 
In Figures~\ref{zzfig10}c,~\ref{zzfig10}d, and~\ref{zzfig10}e, we have overlaid 
the ALMA 1.16 mm continuum emission contours on the {\it JWST} F200W, F356W, and 
F444W images, respectively.  
In the {\it JWST} images ($\lambda$ $>$ 2.0 $\mu$m), embedded point-like sources are seen toward the peaks of the continuum emission (or sources I1 and I2; see cyan contours in Figures~\ref{zzfig10}c,~\ref{zzfig10}d, and~\ref{zzfig10}e), but these IR sources are saturated. 
The positions of the radio continuum sources at 4.9 GHz \citep[from][]{rosero14} are 
found toward these IR sources, and one of these sources might have driven the previously reported SiO(5--4) outflow (see arrows in Figure~\ref{zzfig10}e and also \citet{wang14}). In Figure~\ref{zzfig10}f, one can compare the distribution of the ALMA H$^{13}$CO$^{+}$ and continuum emissions 
with structures/features depicted in the {\it JWST} F444W image.

Taken together these results, we find at least three IR sources embedded in the dense cores (below 8000 AU scale), which are located toward the central part of the envelope-like feature and are associated with the radio continuum emission. Hence, signatures of MSF are evident toward the envelope-like feature. The envelope-like structure is also located at the central hub of G11P1-HFS (see Figures~\ref{fig3lc}a and~\ref{fig3lc}b). 

\begin{figure*}
\includegraphics[width=\textwidth]{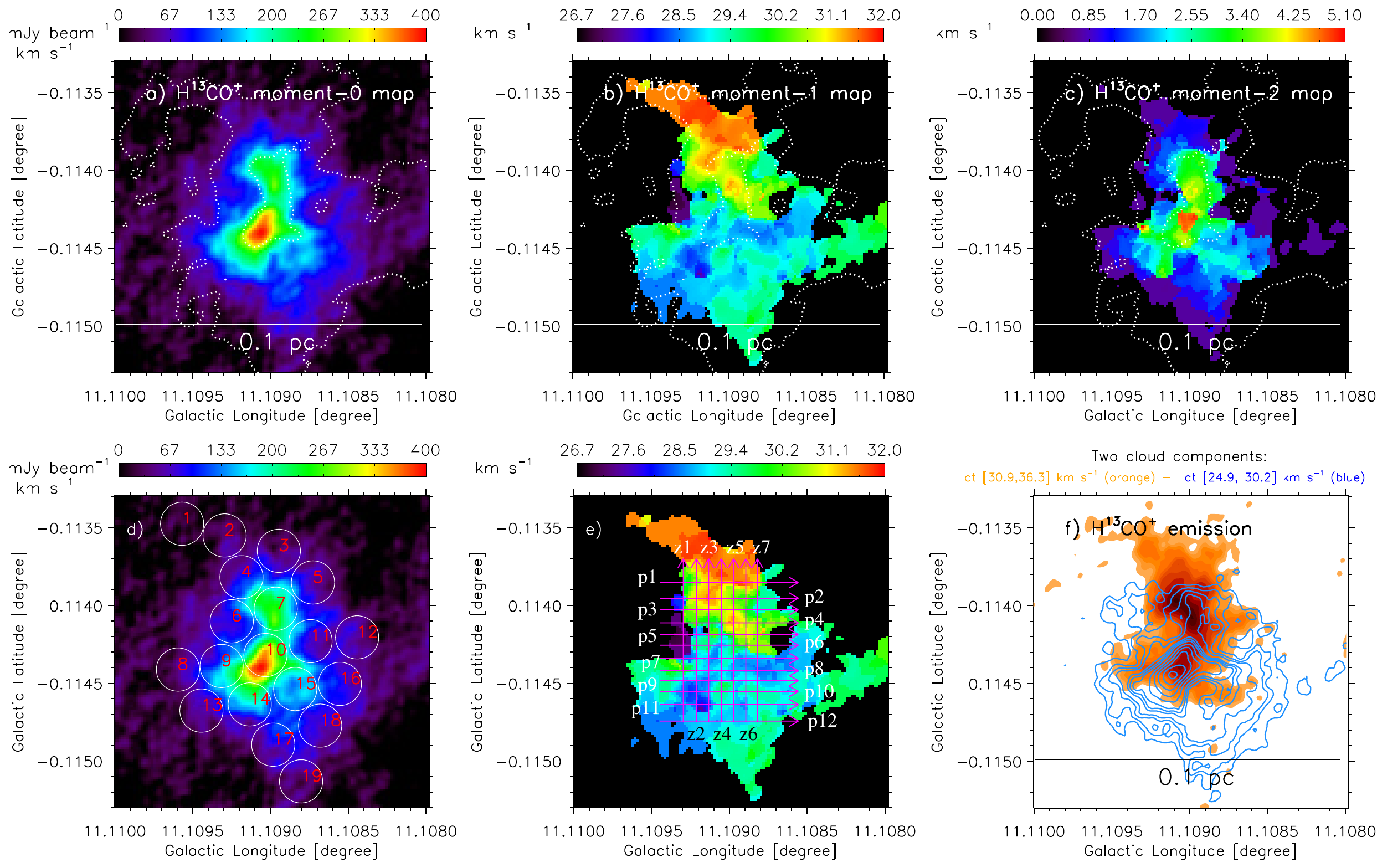}
\caption{Zoomed-in view of an area highlighted by the dot-dashed box in Figure~\ref{fig10}c.
a) ALMA H$^{13}$CO$^{+}$(3--2) moment-0 map at [24.9, 36.3] km s$^{-1}$. 
b) ALMA H$^{13}$CO$^{+}$(3--2) moment-1 map. 
c) ALMA H$^{13}$CO$^{+}$(3--2) moment-2 map.
d) 19 small circular areas (radii = 0\rlap.{$''$}8) are marked on the ALMA H$^{13}$CO$^{+}$(3--2) moment-0 map at [24.9, 36.3] km s$^{-1}$.
e) Several vertical arrows (z1--z7) and horizontal arrows (p1--p12) are highlighted on the ALMA H$^{13}$CO$^{+}$(3--2) moment-1 map. 
f) Distribution of the ALMA H$^{13}$CO$^{+}$(3--2) emission at two different 
velocities (at [24.9, 30.2] and [30.9, 36.3] km s$^{-1}$). Filled contours (in orange; at [30.9, 36.3] km s$^{-1}$) are shown with the 
levels of [0.15, 0.2, 0.25, 0.3, 0.4, 0.45, 0.5, 0.6, 0.7, 0.8, 0.9, 0.98] $\times$ 0.194 
Jy beam$^{-1}$ km s$^{-1}$, while dodger blue contours (at [24.9, 30.2] km s$^{-1}$) are presented with 
the levels of [0.15, 0.2, 0.25, 0.3, 0.4, 0.45, 0.5, 0.6, 0.7, 0.8, 0.9, 0.98] $\times$ 0.287 Jy beam$^{-1}$ km s$^{-1}$. 
In panels ``a'', ``b'', and ``c'', the ALMA continuum dotted contours at [0.03, 0.2] $\times$ 10.4 mJy beam$^{-1}$ are also shown. 
In panels ``a'', ``b'', ``c'' and ``f'', the scale bar shows a spatial scale of 0.1 pc at a distance of 2.92 kpc.} 
\label{fig11}
\end{figure*}

\begin{figure}
\includegraphics[width=9cm]{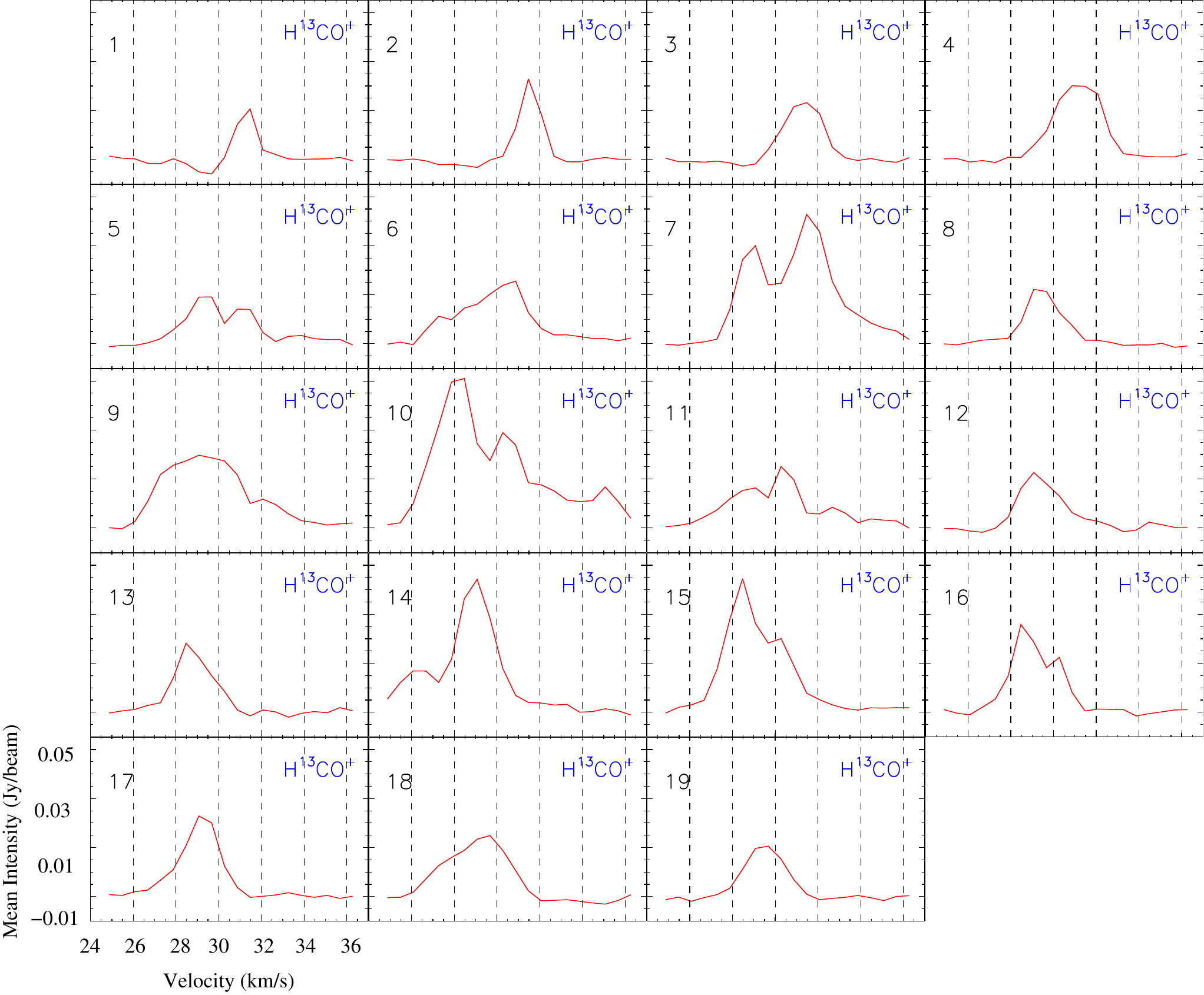}
\caption{H$^{13}$CO$^{+}$(3--2) spectra toward 19 small circular areas (radii = 0\rlap.{$''$}8) as marked 
in Figure~\ref{fig11}d.} 
\label{fig12}
\end{figure}

\begin{figure}
\includegraphics[width=8.3cm]{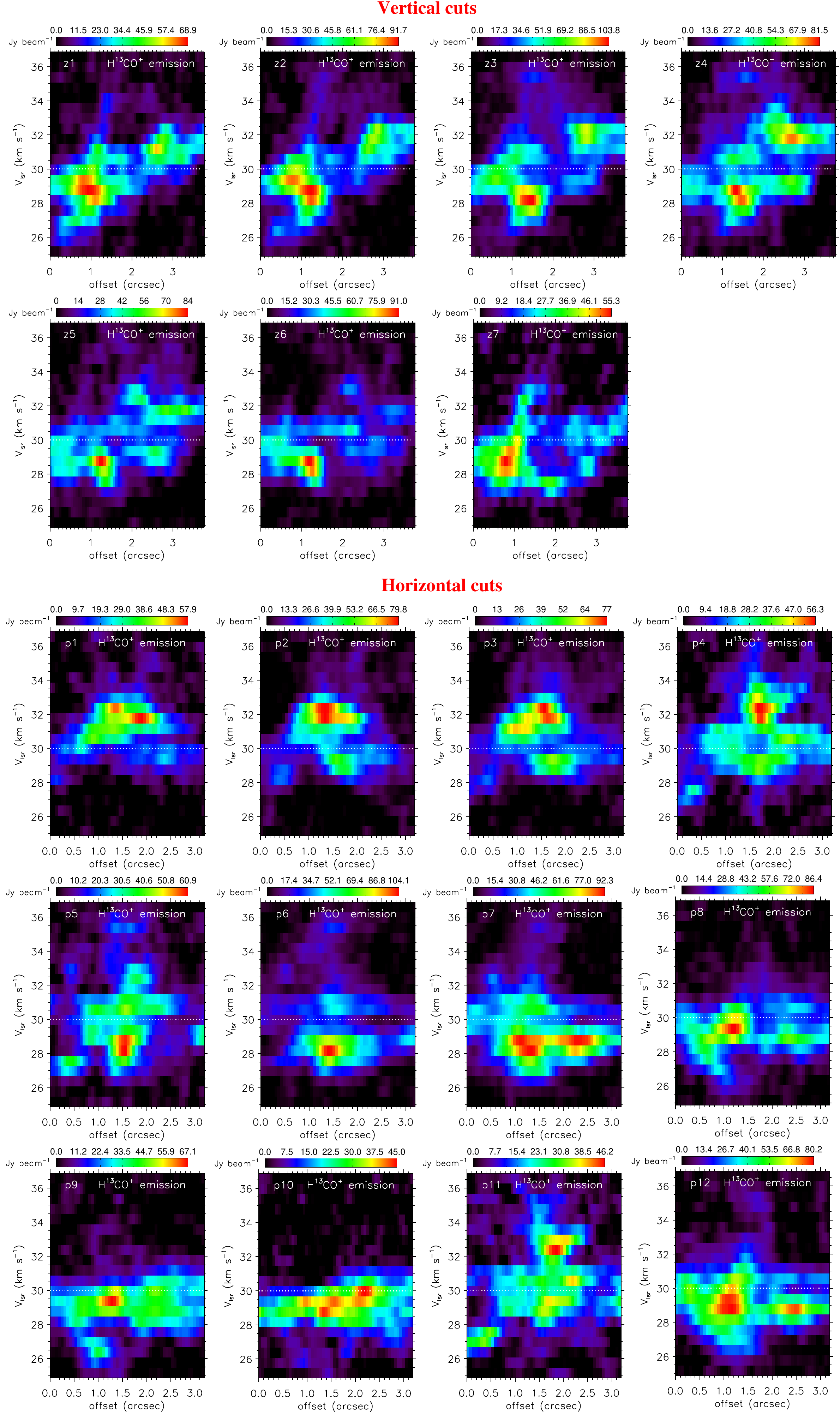}
\caption{Position-velocity diagrams of H$^{13}$CO$^{+}$(3--2) along vertical 
arrows (z1--z7) and horizontal arrows (p1--p12) as marked in Figure~\ref{fig11}e. 
A horizontal dotted line at V$_{lsr}$ = 30 km s$^{-1}$ is also indicated in each panel.} 
\label{fig13}
\end{figure}

To further explore the envelope-like feature (see the dot-dashed box in Figure~\ref{fig10}c), 
we have examined moment maps produced using the ALMA H$^{13}$CO$^{+}$(3--2) line data. 
Figures~\ref{fig11}a,~\ref{fig11}b, and~\ref{fig11}c display the moment-0 map at [24.9, 36.3] km s$^{-1}$, moment-1 map, and moment-2 map 
of an area hosting the envelope-like feature. Higher velocity dispersions 
and a noticeable velocity difference (i.e. 2 km $^{-1}$) are found toward the envelope-like feature. 
In the direction of I1 and I2, the values of velocity dispersion are about 4--5 km s$^{-1}$. 
To further study the molecular gas, 19 small circular areas (radii = 0\rlap.{$''$}8) are indicated in the 
H$^{13}$CO$^{+}$ moment-0 map (see Figure~\ref{fig11}d), where average spectra are produced (see Figure~\ref{fig12}). 
In Figure~\ref{fig11}e, we have selected several vertical arrows (z1--z7) and horizontal arrows (p1--p12) 
that are marked in the ALMA H$^{13}$CO$^{+}$(3--2) moment-1 map, where position-velocity diagrams are generated (see Figure~\ref{fig13}). 
Figure~\ref{fig11}f displays the distribution of the ALMA H$^{13}$CO$^{+}$(3--2) emission at two different 
velocities ranges at [24.9, 30.2] and [30.9, 36.3] km s$^{-1}$, which are chosen based on the 
examination of spectra and position-velocity diagrams of the H$^{13}$CO$^{+}$(3--2) emission (see Figures~\ref{fig12} and~\ref{fig13}). 
The dendrogram structures I1 and I2 are seen at the common zones of these two cloud components. 

Figure~\ref{fig12} presents the spectra of the H$^{13}$CO$^{+}$(3--2) emission toward small circular areas as marked in Figure~\ref{fig11}d. 
Two velocity peaks (around 29 and 31 km s$^{-1}$) are observed toward six small circular areas (i.e., \#5, 7, 10, 11, 15, and 16). 
A single velocity peak is found between 30 and 32 km s$^{-1}$ in the direction of areas \#1, 2, 3, 4, and 6.
In the direction of areas \#8, 9, 12, 13, 14, 17, 18, and 19, another single velocity peak is detected between 28 and 30 km s$^{-1}$.
In Figure~\ref{fig13}, we display the position-velocity diagrams of the H$^{13}$CO$^{+}$(3--2) emission 
toward horizontal and vertical arrows indicated in Figure~\ref{fig11}e. These diagrams also allow us to 
infer two cloud components (see panels z1, z2, z3, z4, p4, p5, and p11 in Figure~\ref{fig13}). 

Overall, two velocity peaks or cloud components with a velocity separation of $\sim$2 km s$^{-1}$ are 
present, as confirmed by our detailed examination of the ALMA line data, towards the envelope-like structure. 
\section{Discussion}
\label{sec:disc}
The selected target G11 does not host any extended H\,{\sc ii} regions, and is not associated with any significant stellar feedback. 
The IRDC G11 contains massive dust clumps and forming massive protostars (see Sections~\ref{sec0} and~\ref{subsec:temp}). 
The surface density map of protostars generated in this work supports the early stages of ongoing star formation activities in the entire IRDC (see Figure~\ref{fig4}d). 
In Section~\ref{ssubsec1a}, at least nine {\it getsf} extracted sources (mass $>$ 65 M$_{\odot}$) are distributed towards G11, and one of the sources associated with G11P1 follows the {\it M-R} criteria of MSF. 
\subsection{Multiple HFS candidates at multi-scale in G11}
\label{aazsec:disc1}
The existence of HFSs in star-forming regions was previously proposed by \citet{myers09}. 
In such configurations, several parsec scale converging interstellar filaments with large aspect ratios and lower column densities surround
the central hub, which has a low aspect ratio and higher column density \citep[e.g.,][]{myers09}. 
After the availability of the {\it Herschel} sub-mm data, the study of the HFSs has attracted a lot of attention. 
In the literature, small-scale HFSs \citep[extent $<$ 6--8 pc;][]{peretto14,dewangan23gl} and 
large-scale HFSs \citep[exent $\sim$10--20 pc; e.g.,][]{morales19,dewangan20,kumar20,bhadari22,mallick23} are commonly detected features in star-forming regions.
It has been investigated that the central hubs of the HFSs are often associated with active star formation including massive stars \citep{schneider12,dewangan15,dewangan17b,dewangan18,dewangan20,dewangna21w42,dewangan23,dewangan21g25,zhou22}.
The incoming material from very large-scales of 1-10 pc may be funnelled along molecular filaments into the hubs, where clusters of protostars and massive stars may form \citep{Tige+2017,Motte+2018,morales19,zhou23}. It seems to promote the longitudinal inflow along filaments \citep[e.g.,][]{vazquez19,padoan20}. 
Hence, HFSs are very important targets to study the mechanism of mass accumulation in MSF. 
In relation to HFSs, an evolutionary scheme (i.e., global non-isotropic collapse (GNIC)) that relies on gravity-driven inflow has been proposed to explain the formation of massive stars \citep{Tige+2017,Motte+2018}. 
According to this scheme, a low-mass protostellar core becomes a high-mass protostar by accumulating mass via gravity-driven inflow, 
which eventually gives rise to an H\,{\sc ii} region powered by a massive OB star. 
Concerning the observed HFSs, \citet{kumar20} also proposed the Filaments to Clusters (F2C) scheme, which includes four stages for MSF \citep[see also][]{beltran22,liu23}.

Overall, the previously published works show that a single HFS, whether it is small-scale or large-scale, is commonly investigated in star-forming regions. Additionally, HFSs are also investigated at the ends of isolated and long filaments undergoing end-dominated collapse (EDC) or edge-collapse \citep[e.g.,][]{bastien83,pon12,clarke15,heigl22}, where high overdensities (or star-forming activities) are observed \citep[e.g.,][]{dewangan19x,Bhadari2020}. One can notice that examples of such configuration are very limited in the literature \citep[see][and references therein]{dewangan23}. Based on the existing literature, how many HFSs are present in a given star-forming site or a filamentary cloud is still a matter of debate.

Our observational findings based on the {\it Spitzer} images show that the cloud G11 is one of the most uncommon sites that is home to multiple IR-dark HFS candidates (HFS1--4; extent $<$ 6 pc; see Figure~\ref{fig3}), where massive clumps and signposts of intense star formation (i.e., outflows, protostars, and masers) are found (see Section~\ref{subsec:temp}).  
One of the selected HFS candidates (i.e., HFS3 or G11P6) is observed by the {\it JWST} facility. High resolution {\it JWST} NIR images also confirm the existence of this HFS candidate (see Section~\ref{nircamsc}), where the presence of embedded sources is evident in the {\it JWST} images.

Interestingly, the {\it JWST} NIR images also discover a small-scale HFS (i.e., G11P1-HFS), which is not resolved in the {\it Spitzer} images (see Figures~\ref{fig3} and~\ref{fig3lc}). One can note that G11 hosts multiple IR-dark HFS candidates at multi-scale.  
Using the ALMA data, we have investigated the dusty envelope-like feature containing cores at the central part of G11P1-HFS and 
its center hosts forming massive stars (below 8000~AU scale; see Section~\ref{linsec2}). However, these ALMA cores are not very massive (see Table~\ref{tab3}).
Previously, using the ALMA 865 $\mu$m map (resolution $\sim$0\rlap.{$''$}3), an analogous configuration has been investigated in the young O-type protostar W42-MME, the infrared counterpart of the 6.7 GHz methanol maser, where a dusty envelope contains continuum sources inner of 9000 AU \citep[see Figure 4d in][]{dewangna21w42}. 
In the case of W42-MME, \citet{dewangna21w42} suggested that prior to core collapse, the massive protostar's core does not build up all of its mass; instead, the core and embedded protostar both grow mass at the same time. This proposal seems to be applicable in G11P1 associated with G11P1-HFS.

The proposed HFS candidates HFS1--4 deserve further investigations using the line and continuum observations at longer wavelengths. 
The birth processes of these HFS candidates are discussed in Section~\ref{zsec:disc1}. 
\subsection{Existence of two velocity components and sub-filaments in G11}
\label{zsec:disc1}
The cloud G11 stands out among other known star-forming regions due to its unique characteristics, which include cloud components, sub-filaments, 
velocity oscillations, embedded protostars, and multiple IR-dark HFS candidates at multi-scale. These features are associated with massive clumps and distinct signatures of star formation (see Section~\ref{sec:data}). 

With the knowledge of radial velocities, the IRDC G11 is divided into two parts, which are part-A around V$_{lsr}$ = 31.5 km s$^{-1}$ and part-B around V$_{lsr}$ = 29.5 km s$^{-1}$. Each part hosts two IR sub-filaments. A noticeable velocity variation toward both the cloud components is found. As mentioned earlier, we find more numbers of the ATLASGAL clumps toward part-B compared to part-A.   

Considering the existence of two cloud components and their overlapping zone, the applicability of cloud-cloud collision scenario may be examined in G11. One of the major observational signposts of cloud-cloud collision is the spatial and velocity connections of two cloud components \citep[e.g.,][]{torii11,torii15,torii17,fukui14,fukui18,fukui21,dhanya21,maity22}. 
Additionally, one expects a bridge feature in position-velocity diagrams and a complementary distribution (i.e., a spatial fit between ``key/intensity-enhancement'' and ``cavity/keyhole/intensity-depression'' features) in the collision event \citep[e.g.,][]{fukui18,dewangan18N36,Enokiya21,maity22,maity23}. 
The two cloud components in the current work seem to be spatially interconnected, and their connection in velocity is also found. 
But, the velocity separation between cloud components toward G11 is only 2 km s$^{-1}$. The analysis of the molecular line data does not favour the presence of any complementary distribution of two clouds in our target area.   
The location of the proposed HFS candidate HFS3 including the massive protostellar candidate G11P1 is clearly visible at the most prominent overlapping zones of the cloud components. 

\citet{nakamura14} used molecular line observations to study the filamentary ridges in the Serpens South IRDC. They found the protocluster clump in the area where these ridges converge. Additionally, they proposed that cluster formation may be caused by the filament-filament collision or the collision of filamentary ridges. Observational results in favour of the filament-filament collision were also reported in recent published works \citep[e.g.,][]{duarte11,henshaw13,frau15,dewangan17a}.
Recently, \citet{dewangan22l} studied an IRDC G333.73+0.37 (d = 2.35 kpc), which hosts previously known two H\,{\sc ii} regions located at its central part. 
Using the {\it Spitzer} 8.0 $\mu$m image, two filamentary structures (length $>$ 6 pc) and a HFS in absorption were investigated toward this IRDC G333.73+0.37.
\citet{dewangan22l} also reported two velocity components (around $-$35.5 and $-$33.5 km s$^{-1}$) toward the IRDC G333.73+0.37, which allowed them to propose a scenario of cloud-cloud collision or converging flows in the IRDC to explain star formation activities. 
In the F2C paradigm of star formation, \citet{kumar20} proposed that the density-enhanced hub is developed when flow driven filaments collide. 
A relatively quiescent stage of star formation is anticipated in such a hub. This particular mode appears to be referring to the IR-dark HFSs \citep[e.g.,][]{liu23}. 
Apart from the observational works, the results of smoothed particle hydrodynamics simulation carried out by \citet{balfour15} demonstrated the formation of hub-filament structure in the process of cloud-cloud collision. Depending on the collision velocity, the shock-compressed layer produced by the collision event can disintegrate into a network of filaments or an array of predominantly radial filaments. 
\citet{beltran22} utilized the N$_{2}$H$^{+}$(1--0) observations to explain the observed HFS and 
massive protocluster in the G31.41+0.31 cloud by a collision event. They also pointed out that the HFS G31.41+0.31 can serve as a reference point for the F2C paradigm of star formation.

Most recently, \citet{dewangan23} reported the presence of intertwined sub-filaments in the dust and molecular maps in the direction of IC 5146 Streamer, and one HFS was also found toward both the edges of the Streamer. 
They suggested the applicability of the ``fray and fragment'' scenario \citep{tafalla15,clarke17} to explain the intertwined sub-structures in IC 5146 Streamer. 
According to the scenario, the primary filament will first form as a result of the collision of two supersonic turbulent gas flows, and that self-gravity and residual turbulent motions will then support the formation of an interconnected system of velocity-coherent sub-structures within the primary filament \citep[see also][]{smith14,shimajiri19}.

The observed velocity oscillation may be the outcome of the core/fragment formation and the large-scale physical oscillation along the filament \citep[e.g.,][]{liu19} and the presence of sub-filaments \citep[e.g.,][]{dewangan2021}.

Considering all the observed results at multi-scale and the highlighted scenarios, multiple physical processes seem to be applicable to the cloud G11, which include the collision process, the ``fray and fragment'' mechanism, and the ``GNIC'' scenario seem to be applicable to the cloud G11. In this relation, it is possible to employ gravity- and turbulence-driven processes together in G11.
\section{Summary and Conclusions}
\label{sec:conc}
A careful inspection of the IR images has allowed us to initiate the present work on the `Snake' nebula or 
IRDC G11.11$-$0.12 or G11 (distance $\sim$2.92 kpc; length $\sim$27 pc), which focuses on unravelling the ongoing star 
formation processes in the IRDC G11. The IRDC G11 has been proposed as an active star-forming site, 
and contains massive dust clumps and forming massive protostars.
The IRDC does not host any extended H\,{\sc ii} regions. 

{\it Spitzer} images suggest the presence of sub-filaments (in absorption) and four IR-dark HFS candidates (extent $<$ 6 pc; HFS1--4) toward G11.
The {\it Herschel} column density and temperature maps also support the existence of the sub-filaments (in emission). 
The $^{13}$CO(2--1), C$^{18}$O(2--1), and NH$_{3}$(1,1) line data reveal that the entire cloud G11 consists of two distinct 
cloud components (at [30.5, 34] and [26, 30.25] km s$^{-1}$). 
The left part (or part-A) of G11 is traced at [30.5, 34] km s$^{-1}$, while the right part (or part-B) of G11 is depicted at [26, 30.25] km s$^{-1}$.
The overlapping area of these cloud components is identified toward the central part of G11, which harbours one HFS (i.e., HFS3). 
Both these cloud components house two IR sub-filaments. 
The molecular line data also show a noticeable velocity oscillation toward G11.  
On the basis of the distribution of the ATLASGAL dust continuum clumps at 870 $\mu$m, 
we find more ATLASGAL clumps toward part-B in comparison to part-A. 
G11 also contains at least nine {\it getsf} extracted sources (mass $>$ 65 M$_{\odot}$), and one of the sources associated with 
the massive protostellar candidate G11P1 satisfies the MSF requirements. 

The \emph{JWST} NIR images discover one IR-dark HFS candidate (extent $\sim$0.55 pc) toward the previously known massive protostar G11P1 (i.e., G11P1-HFS). 
Furthermore, in the direction of HFS4 or G11P6, the existence of an IR-dark HFS candidate is also confirmed by the \emph{JWST} NIR images.  
Combining the results from {\it Spitzer} and \emph{JWST} images, the existence of multiple infrared-dark HFS candidates in G11 is evident, 
where massive clumps ($>$ 500 {\it M$_{\odot}$}) and protostars are distributed. 

The ALMA 1.16 mm continuum map and H$^{13}$CO$^{+}$(3--2) 
line data are employed to study the inner environment of G11P1 associated with G11P1-HFS. 
The ALMA continuum map shows multiple finger-like features (extent $\sim$3500--10000 AU) surrounding a dusty envelope-like feature (extent $\sim$18000 AU) 
toward the central hub of G11P1-HFS. Signatures of forming massive stars are found toward the center of the envelope-like feature, 
where embedded NIR sources associated with radio continuum emission are located ($<$ 8000 AU scale). 
Hierarchical structures in G11P1 are inferred in the ALMA continuum map and the \emph{JWST} NIR images.   
In the direction of G11P1, two cloud components with a velocity separation of $\sim$2 km s$^{-1}$ are investigated in the ALMA line data.
This finding at small-scale (1000--20000 AU) seems to be related with the outcomes derived using the molecular emission at large-scale ($>$ 1 pc). 

Considering our derived outcomes, we propose the applicability of the collision 
of filamentary clouds (mass $>$ $10^4$ {\it M$_{\odot}$}), the ``fray and fragment'' mechanism, and the ``global non-isotropic collapse'' scenario toward the Galactic `Snake' nebula G11, which may explain the ongoing star formation activities and the observed morphologies. 

\section*{Acknowledgments}
We thank the reviewer for useful comments and suggestions, which greatly improved this manuscript. 
The research work at Physical Research Laboratory is funded by the Department of Space, Government of India. 
We acknowledge A.~Men'shchikov for employing the {\it getsf} and {\it hires} on the {\it Herschel} images as well as for carefully reading the draft and providing valuable feedback. This work is based [in part] on observations made with the {\it Spitzer} Space Telescope, which is operated by the Jet Propulsion Laboratory, California Institute of Technology under a contract with NASA. This publication is based on data acquired with the Atacama Pathfinder Experiment (APEX) under programmes 092.F-9315 and 193.C-0584. APEX is a collaboration among the Max-Planck-Institut fur Radioastronomie, the European Southern Observatory, and the Onsala Space Observatory. The processed data products are available from the SEDIGISM survey database, which was constructed by James Urquhart and hosted by the Max Planck Institute for Radio Astronomy. This paper makes use of the following ALMA data: ADS/JAO.ALMA\#2017.1.00101.S. ALMA is a partnership of ESO (representing its member states), NSF (USA) and NINS (Japan), together with NRC (Canada), MOST and ASIAA (Taiwan), and KASI (Republic of Korea), in cooperation with the Republic of Chile. The Joint ALMA Observatory is operated by ESO, AUI/NRAO and NAOJ. The National Radio Astronomy Observatory is a facility of the National Science Foundation operated under cooperative agreement by Associated Universities, Inc. This research made use of astrodendro, a Python package to compute dendrograms of Astronomical data\footnote[1]{http://www.dendrograms.org}. This publication makes use of molecular line data from the Radio Ammonia Mid-Plane Survey (RAMPS). RAMPS is supported by the National Science Foundation under grant AST-1616635. This work is based [in part] on observations made with the NASA/ESA/CSA James Webb Space Telescope. The data were obtained from the Mikulski Archive for Space Telescopes at the Space Telescope Science Institute, which is operated by the Association of Universities for Research in Astronomy, Inc., under NASA contract NAS 5-03127 for \emph{JWST}. These observations are associated with the program \#1182\footnote[2]{http://archive.stsci.edu/doi/resolve/resolve.html?doi=10.17909/08dt-2229}. C.E. acknowledges the financial support from grant RJF/2020/000071 as a part of the Ramanujan Fellowship awarded by the Science and Engineering Research Board (SERB), Department of Science and Technology (DST), Govt. of India. 
\subsection*{Data availability}

The NVSS 1.4 GHz continuum data underlying this article are available from the publicly accessible website\footnote[3]{https://www.cv.nrao.edu/nvss/postage.shtml}. 
The RAMPS NH$_{3}$ (1, 1) line data underlying this article are available from the publicly accessible website\footnote[4]{https://greenbankobservatory.org/science/gbt-surveys/ramps/ramps-data/}.
The ALMA continuum data underlying this article are available from the publicly accessible JVO ALMA FITS archive\footnote[5]{http://jvo.nao.ac.jp/portal/alma/archive.do/}. 
The ATLASGAL data underlying this article are available from the publicly accessible website\footnote[6]{http://atlasgal.mpifr-bonn.mpg.de/cgi-bin/ATLASGAL\_DATABASE.cgi}. 
The SEDIGISM molecular line data underlying this article are available from the publicly accessible website\footnote[7]{https://sedigism.mpifr-bonn.mpg.de/index.html}. 
The {\it Herschel} and {\it Spitzer} data underlying this article are available from the publicly accessible NASA/IPAC IR science archive\footnote[8]{https://irsa.ipac.caltech.edu/frontpage/}.
The \emph{JWST} NIRCam images underlying this article are available from the MAST archive\footnote[9]{https://mast.stsci.edu/portal/Mashup/Clients/Mast/Portal.html}.
\begin{table*}
\setlength{\tabcolsep}{0.1in}
\centering
\caption{Physical parameters of ALMA 1.16 mm continuum sources (see Figure~\ref{fig10}b). The table contains IDs, 
positions, flux densities, deconvolved FWHM$_{x}$ \& FWHM$_{y}$, and masses of 
the continuum sources estimated at different temperatures.} 
\label{tab3}
\begin{tabular}{lccccccccccccc}
\hline 										    	        			      
  Source label                  & Total Flux &FWHM$_{x}$ $\times$ FWHM$_{y}$ & Mass ({\it M$_{\odot}$})& Mass ({\it M$_{\odot}$})& Mass ({\it M$_{\odot}$})   \\ 
                         &     (mJy)  & ($''$ $\times$ $''$)          &  at $T_d$ = 10 K &  at $T_d$ = 15 K&  at $T_d$ = 25 K       \\ 
\hline 
A  &   274.11  &    14.5             &  95.5 & 50.0 & 25.0 \\
B  &   145.68  &    1.5 $\times$ 2.8 &  50.8 & 26.6 & 13.3 \\
C  &    33.57  &    1.8 $\times$ 1.7 &  11.7 &  6.1 &  3.1 \\
D  &     8.32  &    0.8 $\times$ 0.8 &   2.9 &  1.5 &  0.8 \\
b1 &    41.89  &    0.9 $\times$ 0.8 &  14.6 &  7.6 &  3.8 \\
b2 &    81.31  &    1.7 $\times$ 1.7 &  28.3 & 14.8 &  7.4 \\
b3 &    22.48  &    1.3 $\times$ 1.2 &   7.8 &  4.1 &  2.0 \\
\hline          		
\end{tabular}			
\end{table*}			
%


\bibliographystyle{mnras}
\bibliography{reference} 



\end{document}